\documentclass[a4paper,11pt]{article}

\usepackage{amssymb,amsmath,pstricks,epsfig}
\newcommand{\esim}{\underset{\epsilon\to 0}{\sim}}
\newcommand{\re}{\mathrm{Re\,}}
\newcommand{\cor}[1]{\left\langle #1 \right\rangle}
\newcommand{\bi}[4]{{}^{#4\!\!} #1_{#2}^{\phantom{#2}\! #3}}
\newcommand{\tri}[4]{#1_{#2 #3}^{\phantom{#2 #3}\! #4}}

\title{Conformal perturbation theory beyond the leading order}

\author{
Matthias R.\ Gaberdiel$^{1}$\thanks{\tt E-mail: gaberdiel@itp.phys.ethz.ch}, %
Anatoly Konechny$^{2,3}$\thanks{\tt E-mail: anatolyk@ma.hw.ac.uk} , %
and Cornelius Schmidt-Colinet$^{1}$\thanks{\tt E-mail: schmidtc@itp.phys.ethz.ch}\\
\\ \\
${}^{1}$\textit{Institut f\"ur Theoretische Physik, ETH Z\"urich}\\
\textit{CH-8093 Zurich, Switzerland}\\ \\
${}^{2}$\textit{Department of Mathematics, Heriot-Watt University}\\
\textit{Riccarton, Edinburgh, EH14 4AS, UK}\\ \\
${}^{3}$\textit{Maxwell Institute for Mathematical Sciences, 
Edinburgh, UK}\\
}

\date{\today}  

\begin{document}
\maketitle
\begin{abstract}
Higher order conformal perturbation theory is studied for
theories with and without boundaries. We identify systematically the universal 
quantities in the beta function equations, and we give explicit formulae for 
the universal coefficients at next-to-leading order 
in terms of integrated correlation functions. As an example, we analyse the
radius-dependence
of the conformal dimension of some boundary operators 
for the case of a single Neumann brane on a circle, and for an intersecting brane
configuration on a torus, reproducing in both cases the expected
geometrical answer.

\end{abstract}

\newpage

\section{Introduction}
\renewcommand{\theequation}{\arabic{section}.\arabic{equation}}
\setcounter{equation}{0}

Perturbations of conformal field theories by relevant operators
have been intensively studied starting with the work of 
Zamolodchikov \cite{Zamolodchikov:1987ti,Zamolodchikov:1989zs}
on integrable perturbations of conformal field theories. 
Numerous examples have been considered, but there are also 
a number of structural results, in particular 
the $c$-theorem of Zamolodchikov \cite{Zamolodchikov:1986gt}
that states that the central charge $c$ cannot increase along
renormalisation group flows, as well as the analogous $g$-theorem 
\cite{Friedan:2003yc} for the boundary entropy \cite{Affleck:1991tk}.
Perturbations of conformal field theories also play an important
role in string theory, for example for time dependent
backgrounds, see {\it e.g.}\ \cite{Freedman:2005wx,Graham:2006gca}. 

In the context of string theory also {\it marginal} perturbations
are of significance. Most string theories of interest possess moduli,
{\it i.e.}\ free parameters such as the size and shape of the
background or the position of some D-brane, and these correspond
to marginal operators in the two-dimensional world-sheet theory. 
Exact conformal field theory solutions are often only available
at special points (in particular the rational points) in moduli space,
and it is important to learn to control the theory away from these special 
points, {\it i.e.}\ after perturbations by marginal operators. 

Usually one thinks of the moduli as corresponding to {\it exactly
marginal} operators, and then the renormalisation group analysis is, 
by definition, trivial. However, in the context of bulk and boundary
perturbations, the situation can be more subtle. In particular, 
exactly marginal bulk operators (describing moduli of the 
closed string background) can cease to be exactly marginal
in the presence of a boundary. If this is the case, they induce a 
non-trivial renormalisation group flow on the boundary \cite{FGK}. 
\smallskip

In the analysis of perturbations by relevant operators, a first
order analysis is usually sufficient (see {\it e.g.}\ \cite{Z2,RRS}). 
However, in the context of
marginal perturbations, it is often necessary to go to higher
order in perturbation theory. The simplest example of such a situation
is a single Neumann brane on a circle, for which the conformal
dimension of the momentum eigenstates depends on the radius
modulus. From the point of view of the world-sheet, the change in conformal 
dimension for this boundary operator does not arise at first
order, but only appears at next-to-leading order 
in perturbation theory. 

Higher order conformal perturbation theory also plays a role in
proofs of integrability of particular 
bulk and/or boundary perturbations \cite{Zamolodchikov:1989zs,CL,GZ}.
Conformal perturbation theory at higher orders was  
studied  for particular models in \cite{Lassig,CF,CCF,Ludwig:2002fu};
general aspects of the pure bulk case were also discussed in 
\cite{GM1,GM2}.
\smallskip

In this paper we make an attempt at a systematic analysis of conformal 
perturbation theory beyond the leading order. We begin by
analysing  which RG coefficients are scheme-independent
(or universal) and thus can have a physical interpretation. (In particular, 
we show that this is the case for the  coefficient describing the change
in conformal dimension of the momentum fields on the Neumann brane.)
We then outline a specific scheme --- the position space minimal
subtraction scheme --- in which higher order RG coefficients
can be calculated. This allows us to prove that the combined 
bulk-boundary perturbation problem is renormalisable at the 
quadratic order. While the minimal subtraction
scheme is conceptually clean, explicit calculations of the 
RG coefficients are often rather cumbersome. We therefore
also consider another, Wilsonian type  scheme, 
to which we refer to as the `OPE scheme` since the first nontrivial 
terms in the beta functions are given by OPE coefficients.
This scheme has some conceptual shortcomings at higher orders but 
is computationally somewhat simpler. For the universal coefficients we 
are interested in, the result is independent of which of  the two 
schemes we use (as we also verify explicitly). We can
therefore determine the coefficients of interest (in particular
the formula for the shift in conformal dimension for the momentum
fields on the Neumann brane) in the Wilsonian
approach. In the resulting formulae the universal quantities are expressed as  
integrals over certain correlation functions.
As an illustration  we also apply the formulae to an intersecting brane model 
on a torus, and  again reproduce the geometric result.
\medskip

The paper is organised as follows. In section~2 we discuss which
RG coefficients in the boundary beta function are universal in the 
presence of marginal bulk perturbations 
 (section~2.1). We then introduce the minimal subtraction 
scheme, both for pure boundary perturbations (section~2.2.1), as 
well as for the combined bulk-boundary problem for which
we prove renormalisability at the quadratic order  (section~2.2.2).
We also introduce the Wilsonian scheme and discuss its advantages 
and shortcomings (section 2.3). Finally, we explain how the 
discussion can be generalised to include boundary changing operators
(section~2.4). In section~3 these ideas are applied to two 
examples, the single Neumann brane on a circle (section~3.1), as well as a
configuration of two intersecting D1-branes on a 2-torus (section~3.2).
Finally, we discuss in section~4
how our techniques for the calculation of higher order RG coefficients
can also be applied to pure bulk or pure boundary perturbation theory.

\section{Bulk-boundary  perturbations of  BCFTs}
\setcounter{equation}{0}
Let us  start with a general discussion of the effect of perturbations by  
marginal bulk operators on boundary degrees of freedom. Generically 
such a deformation 
will induce a renormalisation group (RG) flow on the space of 
boundary conditions  \cite{FGK}. Under a certain condition, to be 
formulated precisely below, the induced boundary deformation is
however scale independent to the first order in the bulk deformation 
parameter.  In this case one can study how the set of boundary scaling 
dimensions changes with the bulk deformation. We will demonstrate that 
(despite no occurrence of RG flows)  the RG technique is very useful in 
addressing  this question. In particular we will derive, using the RG methods,
general expressions for a first order change in  dimensions 
of  boundary operators along a bulk deformation.

\subsection{Universal terms in marginal bulk perturbations}\label{sec2.0}

Before we give a detailed discussion we need to introduce some notation.
Consider a  boundary conformal field theory (BCFT) defined on 
the upper half plane $\mathbb{H}^+=\{(x,y)|y\ge 0\}$ with complex 
coordinate $z=x+iy$. Let  $\phi_k(z,\bar{z})$ be bulk primary fields with 
conformal weights $(h_k,h_k)$ so that their scaling dimensions 
are $\Delta_k=2h_k$. For a single (fundamental) conformal boundary condition  
we denote the boundary primaries by $\psi_p(x)$, and their scaling
dimensions by $h_{p}$. Later we will generalise our discussion to
superpositions of conformal boundary conditions. 
We will assume that the two point functions are normalised 
as
\begin{equation}\label{two-point_normalisation}
\langle \phi_i(z,\bar z) \phi_j(w,\bar w)\rangle = 
\frac{\delta_{ij}}{|z-w|^{2\Delta_{i}}} \ , \quad 
\langle \psi_p(x)\psi_q(y) \rangle =
\frac{\delta_{pq}}{|z-w|^{2h_{p}}}  \ .
\end{equation}
In particular, this means that we assume all fields to be self-conjugate;
this is obviously not a real restriction, and our analysis can easily be
generalised. The operator product expansion (OPE) for pairs of bulk and 
boundary operators has the form 
 \begin{eqnarray}
\phi_i(z,\bar{z})\phi_j(w,\bar{w}) & = & \sum_{k}C_{ij}{}^{k}
|z-w|^{\Delta_k-\Delta_i-\Delta_j}\phi_k(w,\bar{w})+\ldots\,, 
\label{bulkbulkOPE} \\
\psi_p(x)\psi_q(y) & = & \sum_{r}D_{pq}{}^{r}
(y-x)^{h_r-h_p-h_q}\psi_r(y)+\ldots \quad (y>x) \ . \label{boundaryboundaryOPE}
\end{eqnarray}
Finally, when a bulk operator approaches the  boundary  it can be expanded 
using the bulk to boundary OPE 
\begin{equation}\label{bulkboundaryOPE}
\phi_k(x+iy,x-iy)=\sum_{p}B_{k}{}^{p}
(2y)^{h_p-\Delta_k}\psi_{p}+\ldots\ . 
\end{equation}
\smallskip

With these preparations, let us now consider a perturbation  of the given 
BCFT generated by the Euclidean action  perturbation 
\begin{equation}\label{deltaS}
\delta S=\sum_k l^{\Delta_k-2}\lambda^k\iint\!\! dxdy\,\phi_k(x,y)
+\sum_{p}l^{h_p-1}\mu^{p} \int\!\! dx\,\psi_p(x)\  .
\end{equation}
Here $\lambda^k$, $\mu^{p}$ are the dimensionless  coupling constants 
of the respective operators, and $l$ is a renormalisation  distance  scale. 
Up to second order in the coupling constants, 
the beta functions have the following general form
\begin{eqnarray}
\beta^k&=&y_k\lambda^k+\sum_{ij}\mathcal{C}_{ij}^{k}
\lambda^i\lambda^j + \ldots 
\label{genblkbeta}\\[5pt]
\beta^{p}&=&y_p\mu^{p} + \sum_i{\mathcal{B}}_{i}^{p}
\, \lambda^i+\sum_{qr}\mathcal{D}_{qr}^{p}\mu^{q}
\mu^{r}+\sum_{iq}\mathcal{E}_{iq}^{p}\lambda^i\mu^{q}+\ldots \ .
\label{genbdrybeta}
\end{eqnarray}
Here, $y_k=2-\Delta_k$, and $y_q=1-h_{q}$ are the bulk and 
boundary anomalous dimensions, respectively.
The omitted terms stand for higher orders in the coupling constants.
A general property of any local RG scheme is that the bulk beta functions are 
independent of the boundary couplings.
\smallskip

It has been known for quite some time \cite{Polyakov}
(see also \cite{Cardy}) that in a particular renormalisation scheme 
the coefficients $\mathcal{C}_{ij}^{k}$ for  bulk theories 
are given by $\mathcal{C}_{ij}^{k}=\pi \tri{C}{i}{j}{k}$, 
where $C_{ij}{}^{k}$ are the bulk OPE coefficients from (\ref{bulkbulkOPE}).
The same scheme, to be discussed in more detail in section 2.3, can be easily adapted 
for theories on the half plane. The coefficients $\mathcal{D}_{qr}^{p}$ 
coincide then with the boundary structure constants 
$D_{qr}{}^{p}$ (see {\it e.g.}\ \cite{AL}), and for the coefficients ${\mathcal{B}}_{i}^{p}$ 
we have  ${\mathcal{B}}_{i}^{p}=\frac{1}{2}B_{i}{}^{p}$, where $B_{i}{}^{p}$ 
are the bulk to boundary OPE coefficients from 
(\ref{bulkboundaryOPE}), see \cite{FGK}.

Consider now the case  where we perturb the BCFT by a single bulk field 
$\phi(x,y)$  with a coupling constant $\lambda$. Furthermore we want to assume 
that the bulk beta function $\beta^{\phi}(\lambda)$ vanishes. 
In this case, even in the absence of  an initial boundary 
perturbation $\mu_{\rm bare}^{p}=0$, a boundary renormalisation group flow can
be triggered by the terms ${\mathcal{B}}_{\phi}^{p}\lambda$ in the boundary 
beta function.  Such boundary terms, however, are in general not universal. For
example, if the induced boundary fields are all relevant, {\it i.e.}\ $y_p>0$, then the 
corresponding terms in the boundary beta function can be removed by a coupling 
constant redefinition 
\begin{equation}\label{redef}
\mu^{p}\mapsto \tilde \mu^{p} = \mu^p + \frac{{\mathcal{B}}^{p}_{\phi}}{y_p}\lambda \ .
\end{equation}
The above coupling constant redefinition looks peculiar in that $\tilde \mu^p$ is 
not proportional to $\mu^p$. It has, however, a simple meaning. 
Let  $Z=\langle e^{\delta S}\rangle $ be the renormalised partition 
function\footnote{Here we are talking about the partition function on a disc. 
The passage from the half plane to the disc is straightforward because the bulk 
theory is conformal.} of 
the perturbed theory (\ref{deltaS}). 
We have  
\begin{equation} \label{Zredef}
\left(\frac{\partial \ln Z}{\partial \lambda}\right)_{\{\tilde\mu^p\}} -
\left(\frac{\partial \ln Z}{\partial \lambda}\right)_{\{ \mu^p\}} = -
\sum_p \frac{{\mathcal{B}}_{\phi}^{p}}{y_p}
\left(\frac{\partial \ln Z}{\partial \tilde \mu^p}\right)_{\lambda} = -
\sum_p \frac{{\mathcal B}_{\phi}^{p}}{y_p}
\left(\frac{\partial \ln Z}{\partial  \mu^p}\right)_{\lambda} \ ,
\end{equation}
where the partial derivatives on the left hand side are taken with 
the boundary constants $\mu^{p}$ or $\tilde \mu^p$ held fixed.
The identities (\ref{Zredef}) mean that, after the redefinition (\ref{redef}),
the bulk coupling constant $\lambda$ couples to a re-defined field 
\begin{equation}\label{redef2}
 \tilde \phi(x,y) = \phi(x,y) -
\sum_p \frac{{\mathcal{B}}_{\phi}^{p}}{y_p}\, \psi_{p}(x)\, \delta(y) \  .
\end{equation}
Part of the renormalisation procedure amounts to defining 
the operator coupling to $\lambda$ on the half plane so that 
the correlation functions involving that operator are distributions 
(and thus integrable in any bounded region on the half plane). 
In the interior of the half-plane the resulting operator must coincide
with the bulk operator $\phi(x,y)$, but in general extra subtractions
may be required at the boundary. 
The redefinition  (\ref{redef2}) stemming from the change of scheme 
(\ref{redef}) reflects  the natural ambiguity in defining such a fully 
subtracted operator extending $\phi(x,y)$.
\smallskip

Suppose now that all terms linear in $\lambda$ can be removed in this 
manner. Then the resulting boundary beta functions have the form
\begin{equation}
\tilde \beta^{p}= \sum_q D_{q}^{p}(\lambda)\tilde \mu^q + {\cal O}(\tilde \mu^2) \ ,
\end{equation}
where 
\begin{equation}
D_{q}^{p}(\lambda) = y_p\, \delta_{q}^{p} + \lambda \, \tilde {\mathcal{E}}_{\phi q}^{p}
 + {\cal O}(\lambda^2) \qquad \hbox{with} \qquad
\tilde  {\mathcal{E}}_{\phi q}^{p}= {\mathcal{E}}_{\phi q}^{p} - 
2\sum_r {\mathcal{D}}^p_{(qr)} \frac{{\mathcal B}_{\phi}^{r}}{y_r} 
\end{equation}
and $\mathcal{D}^p_{(qr)} =\tfrac{1}{2} (\mathcal{D}^p_{qr} + \mathcal{D}^p_{rq})$. 
Now that the boundary beta functions are all proportional  to the boundary 
coupling constants we can treat the boundary perturbations 
(at least those of the relevant operators) infinitesimally to read off the 
dimensions of the boundary operators in the deformed theory.  More 
specifically, we claim that the eigenvalues of the matrix 
$D_{q}^{p}(\lambda)$ are to be identified with $y=1-h$, where $h$ is the
scaling dimension of the boundary operator in the deformed theory, 
and the corresponding boundary primaries are the 
eigenvectors of  $D_{q}^{p}(\lambda)$. To leading order in $\lambda$,
the matrix $D_{q}^{p}(\lambda)$ can be diagonalised by the transformation 
\begin{equation}
\tilde \mu^{p}\mapsto \sum_q (\delta_q^p + \lambda f_q^p)\, \tilde \mu^q \ , \qquad
\hbox{where} \qquad
f_{q}^{p}=\left\{ \begin{array}{cl}
 \frac{\tilde {\mathcal{E}}_{\phi q}^{p}}{y_p-y_q} & \mbox{for $p\ne q$} \\
 0 & \mbox{for $p=q$.}  
 \end{array} \right. 
 \end{equation}
The corresponding primary fields are\footnote{The fields 
$\psi_{p}[\lambda]$ are defined up to adding a multiple of $\lambda \psi_p$.}
\begin{equation}
\psi_{p}[\lambda]= \psi_p - \lambda 
\sum_{q\ne p}\frac{\tilde {\mathcal{E}}_{\phi p}^{q}}{y_q-y_p} \psi_q \ ,
\end{equation}
and their anomalous dimensions are 
\begin{equation}
y_p[\lambda]=y_p + \lambda \, \tilde {\mathcal{E}}_{\phi p}^{p} \  .
\end{equation}
We further claim that the quantity specifying the dimension shifts
\begin{equation}\label{dim_shift}
\tilde {\mathcal{E}}_{\phi p}^{p}= {\mathcal{E}}_{\phi p}^{p}
-2\sum_r {\mathcal{D}}_{(pr)}^p \frac{{\mathcal{B}}_{\phi}^{r}}{y_r}
\end{equation}
is scheme independent. To see this we consider a coupling constant 
redefinition of the form 
\begin{equation}
\mu^p\mapsto \mu^p + \lambda \, b^p + \sum_{qr} d_{qr}^p \mu^q \mu^r 
+ \sum_q e_q^p \lambda \mu^q + \dots \ .
\end{equation}
Under this redefinition the coefficients in the beta functions (\ref{genbdrybeta})
change as
\begin{eqnarray}\label{coef_shifts}
&& {\mathcal{B}}_\phi^p\mapsto {\mathcal{B}}_\phi^p - b^p y_p \ , \nonumber\\
&& {\mathcal{D}}_{rs}^p \mapsto {\mathcal{D}}_{rs}^p + 
d^p_{(rs)}(y_r + y_s - y_p) \ , \nonumber \\
&& {\mathcal{E}}_{\phi r}^p \mapsto {\mathcal{E}}_{\phi r}^p 
-2 \sum_s {\mathcal{D}}_{rs}^p b^s 
-2\sum_s d_{(rs)}^p b^s (y_r + y_s -y_p)  
+2\sum_s d_{(rs)}^p{\mathcal{B}}_{\phi}^{s} + e_{r}^{p}(y_r-y_p)\  .
\end{eqnarray}
It is straightforward to check that under the transformations (\ref{coef_shifts}) 
the quantity (\ref{dim_shift}) is indeed invariant.


\subsection{Computation in a minimal subtraction scheme}

In the following we want to explain in detail how these coefficients  --- in 
particular (\ref{dim_shift}) --- can be calculated explicitly. We shall first
study this question in a minimal subtraction scheme.
In order to make sense of the formal perturbation series we shall use
a point splitting regularisation. In particular, we require that any two perturbing 
bulk or 
boundary fields do not approach each other closer than a cut-off $\epsilon$, and 
that the perturbing bulk fields only approach the boundary up to a distance
$\epsilon/2$.  Before specialising to the bulk and boundary situation discussed 
in the previous section, let us first discuss some generalities of 
renormalisation. For brevity we consider only boundary perturbations but 
that is inessential for the points we want to make. 

\subsubsection{Generalities of minimal subtraction schemes for boundary perturbations}

Let us consider a perturbed BCFT action 
\begin{equation}\label{L1}
S=S_{\rm BCFT} + \sum_{p}\mu^{p}_{\rm B}\int\!\!dx\, \psi_{p}(x)\ ,
\end{equation}
where $\mu^{p}_{\rm B}$ are the bare coupling constants. 
Let $l$ be an infrared distance scale at which we wish to 
renormalise the theory. In terms of the 
renormalised dimensionless coupling constants $\mu^{p}$,
the same Lagrangian  (\ref{L1}) can be expressed as
 \begin{equation}\label{L2}
S=S_{\rm BCFT} + \sum_{p}l^{-y_p}\mu^{p} \int\!\!dx\, \psi_{p}(x) 
+ S_{\rm ct}\ ,  
\end{equation}
where $y_p$ are anomalous dimensions of the fields $\psi_{p}(x)$, and $S_{\rm ct}$ is 
a counterterm action. 
Perturbation theory generates integrals of the form
\begin{equation}\label{pert_gen}
\int\!\!\dots\!\! \int\!\! dx_1\dots dx_n\,  \psi_{p_1}(x_1)\psi_{p_2}(x_2)\dots 
\psi_{p_n}(x_n)\prod_{i<j}^{n} \theta(|x_i-x_j|-\epsilon) \, 
\theta(L-|x_i-x_j|) \ ,
\end{equation}
where we have also introduced an infrared regulator $L$.
The above  expression is to be understood in the operator sense, 
{\it i.e.}\ inside a correlator with arbitrary other insertions. 
The product of fields in (\ref{pert_gen}) can be 
expanded in terms of a complete set of local operators $\Psi_{A}$
as
\begin{equation}\label{multOPE} 
\psi_{p_1}(x_1)\psi_{p_2}(x_2)\dots \psi_{p_n}(x_n) = \sum_{A} 
C^{A}_{p_1,\dots, p_n}(x_1, \dots x_{n-1}) \, \Psi_{A}(x_{n}) \ , 
\end{equation}
where we have arbitrarily chosen the point of insertion on the right hand side 
to be $x_n$.
If the OPEs of the conformal families of the primaries $\psi_{p}$ close on 
themselves, we can take for $\Psi_{A}$ the fields $\psi_{p}$ and their conformal 
descendants. In conformal field theory the expansion  (\ref{multOPE}) always 
converges \cite{Mack} unlike in massive QFTs for which the OPE may be merely  an 
asymptotic expansion. Substituting (\ref{multOPE}) into (\ref{pert_gen}) we obtain 
expressions of the form
\begin{equation}\label{exprO}
\sum_{A} C_{p_1, \dots , p_n}^{A}(\epsilon, L)\int dx\, \Psi_{A}(x) \ ,
\end{equation}
where 
\begin{equation} \label{ints}
C_{p_1, \dots , p_n}^{A}(\epsilon, L)=\int\!\!\dots\!\! \int\!\! dx_1\dots dx_{n-1}\,
C^{A}_{p_1,\dots, p_n}(x_1, \dots x_{n-1})\prod_{i<j}^{n} \theta(|x_i-x_j|-\epsilon)\,  
\theta(L-|x_i-x_j|) \ . 
\end{equation}
The integrals (\ref{ints}) are finite because of the cut-offs.
In the limit $\epsilon \to 0$ the  coefficients $C_{p_1, \dots , p_n}^{A}(\epsilon, L)$ 
diverge with the divergences coming from the regions of integration in which 
two or more insertion points $x_1, \dots x_n$ collide. In fact, if $k$ operators 
$\psi_{p_1}, \dots, \psi_{p_{k}}$ come close together  (with the other insertions 
bounded away from the point of coincidence) to produce an operator 
$\Psi_{S}$,  the leading divergence has the form 
 \begin{equation}\label{ldivs}
 C \epsilon^{y_{p_{1}}+ \cdots + y_{p_{k}}-y_{S}}\,
 L^{y_{S} + y_{p_{k+1}} + \cdots + y_{p_{n}}  -y_A} \ ,
 \end{equation}
where $C$ is some numerical constant. Here we have assumed that the
resonance condition 
\begin{equation} \label{res}
y_{p_{1}}+ \dots + y_{p_{k}}-y_{S}=0
\end{equation}
does not hold;
otherwise the corresponding divergence is logarithmic. 
The above reasoning follows essentially from dimensional counting, as well as from
the locality of the OPE, ensuring the independence of  the expansion (\ref{multOPE}) 
from  $L$. If we only perturb by marginal or relevant fields, $y_{p_j}\geq 0$, then
divergences can only occur if also $\Psi_S$ is relevant, {\it i.e.}\ $y_S>0$. Assuming
that the OPE is closed, $\Psi_{S}$ is then one of  the perturbing relevant 
or marginal primary fields  $\psi_{p}$. In the `minimal subtraction scheme' we 
are using here, we only introduce counterterms for actually divergent contributions; 
the above reasoning then implies that the scheme closes on itself.

The divergences arising when $k$ operators come together first emerge at order 
$n=k$ in perturbation theory. One expects that they can be canceled by local 
($L$-independent) counterterms. These counterterms then also cancel the 
non-local subdivergences  (\ref{ldivs}) that appear at order $m>k$ in perturbation
theory. Thus we only need to deal with the case when $n=k$, in which 
case $\Psi_S$ and $\Psi_A$ must have a non-trivial two-point function, and hence
$y_S=y_A$. Then the coefficient (\ref{ldivs}) is independent of $L$, and hence
converges when $L\rightarrow \infty$.
Note that the lower order counterterms may also contribute to the $k=n$ divergence 
 when the counterterm insertion from the order $l<n$ comes close together with 
 $n-l$ fields  $\psi_{p_{i}}$.
  The same dimensional reasoning however tells us that 
the final coefficient must again be independent of $L$, and the remaining 
divergence can be canceled by a local counterterm. 

The above discussion should however not be taken to be a 
recursive proof of renormalisability of conformal
perturbation theory.  One problem that needs to be tackled  is the classical problem 
of overlapping divergences; in the case at hand this occurs 
when $k$ points come together with a subset of $l<k$ points coming 
together much faster than the remaining ones. The associativity of the OPE 
in conformal field theory should be the key property ensuring the 
consistency in dealing with overlapping divergences, but we have not
attempted to work this out in detail. However, we will see in the concrete
examples of the next subsections how the above discussion can be made more 
rigorous.  In particular we will prove the renormalisability  of conformal 
perturbation theory at the next-to-leading order using analytic properties of 
conformal blocks. 
\medskip     
 
In order to illustrate these ideas, let us now consider an integral that emerges at 
second order in perturbation theory,
\begin{equation}
\frac{1}{2!}\sum_{p,q}l^{-y_p-y_q}\mu^{p}\mu^{q}\int\!\! dx_1\int\!\!dx_2\, 
\psi_{p}(x_1)\psi_{q}(x_2)\,
\theta(|x_1-x_2|-\epsilon) \, \theta(L-|x_1-x_2|)\ .
\end{equation}
The product of the two boundary fields  can be expanded via the OPE  
(\ref{boundaryboundaryOPE}). Performing one of the integrals, we see
that we get ultraviolet divergences of the form (in the limit $L\rightarrow \infty$)
\begin{equation}\label{Sdiv}
S_{\rm div}^{(2)}=-\frac{1}{2}\sum_{p,q,r}\frac{D_{pq}{}^{r}}{y_p+y_q-y_r}
\left(\frac{\epsilon}{l} \right)^{y_p+y_q-y_r}l^{-y_r}\mu^{p}\mu^{q}\int\!\!dx\, \psi_{r}(x) \ ,
\end{equation}
where the summation runs only over those indices $p,q,r$ for which  
$y_p + y_q -y_r < 0$. In particular, $y_r>0$, and thus only relevant primary fields 
$\psi_r$ contribute. As before, we have also assumed here that there 
are `no resonances', {\it i.e.}\ that none of the 
expressions $y_p + y_q -y_r $ vanishes. Then only power divergences
occur at this order. 

The above divergences can be canceled by adding a minimal action 
counterterm $S_{\rm ct}^{(2)}=-S_{\rm div}^{(2)}$. Equating 
the two expressions (\ref{L1}) and (\ref{L2}) we obtain 
up to second order in the coupling constants 
\begin{equation}\label{muB}
\mu_{\rm B}^{r} = l^{-y_r}\Bigl[   \mu^{r} + 
\frac{1}{2} \sum_{p,q \in I_{r}^{(2)}}
\frac{D_{pq}{}^{r}}{y_p+y_q-y_r}
\left(\frac{\epsilon}{l} \right)^{y_p+y_q-y_r}l^{-y_r}  \mu^p \mu^q\Bigr] \ ,
\end{equation} 
where $I_{r}^{(2)}$ is the set of pairs of indices $(p,q)$ for which 
$y_p + y_q -y_r < 0$.
Differentiating both sides of (\ref{muB}) with respect to $l$ with
fixed $\mu^{r}_{B}$ we obtain 
\begin{equation}
l\frac{d\mu^{r}}{dl}=\beta^{r}(\mu)=y_r\mu^{r} \ .
\end{equation} 
Thus the beta functions are linear in $\mu$. It is easy to see that this
property continues to hold also at higher order in perturbation theory, 
as long as the  divergences are power-like. 

On the other hand, if we have a non-trivial resonance at lowest order,
{\it i.e.}\ if $y_r=y_p + y_q$, then formula (\ref{Sdiv}) takes the form 
(we are assuming for simplicity that there are no other divergences 
at this order)
\begin{equation}\label{ex1}
S_{\rm div}^{(2)}=-\frac{1}{2} 
D_{pq}{}^{r} \ln(\epsilon/l) l^{-y_r}\mu^{q}\mu^{p}\int\!\!dx\, \psi_{r}(x) \ ,
\end{equation}
where we cut off the divergent integral in the infrared region at the
renormalisation scale $l$. Introducing a counterterm  
$S_{\rm ct}^{(2)}=-S_{\rm div}^{(2)}$ we then obtain a beta function for the coupling 
$\mu^r$
\begin{equation}\label{ex2}
\beta^{r}=y_r\mu^r + D_{pq}{}^{r}\mu^{p}\mu^{q} \, . 
\end{equation}
More generally, in the minimal subtraction scheme at hand, the nonlinear 
terms in the beta functions all come from resonances. However, in general
not all resonant terms are universal.

\subsubsection{Minimal subtraction scheme for bulk-boundary perturbations}

After this interlude we now return to the case of interest, namely the description
of the minimal subtraction scheme for bulk-boundary perturbations. In fact, the 
above discussion generalises in a straightforward manner to include 
an additional  perturbation  by a bulk  field. For simplicity of presentation,
we shall assume that the bulk field $\phi(z,\bar z)$ is a spinless 
relevant or marginal primary field of scaling dimension $\Delta=2-y_{\phi}$ 
with $y_{\phi}\geq 0$.  As we shall explain below, the bulk-boundary 
perturbation at the next to leading order in perturbation theory is then renormalisable. 
We will  specialise to the situation where the bulk field is marginal ($y_\phi=0$) later.

At the linear order in the bulk coupling $\lambda$ the divergences 
in perturbation theory only arise from singularities as the bulk field 
approaches the boundary. These are described by the bulk to boundary OPE. 
If there are no boundary fields for which $h_p=\Delta-1$ we have 
power divergences of the form  
(\ref{bulkboundaryOPE}) 
\begin{equation}\label{S1}
S^{(1)}_{\rm div}=- \lambda \, \sum_{p\in I^{(1)}}
\frac{B_{\phi}{}^{p}\, }{2(y_{\phi}-y_p)} \,
\left(\frac{\epsilon}{l} \right)^{y_{\phi}-y_p} \, l^{-y_p} 
\int\!\! dx \, \psi_{p}(x) \ ,
\end{equation}
where 
\begin{equation}
I^{(1)}=\{ p\, |\, y_{p}>y_{\phi}\}\, .
\end{equation}
In the minimal subtraction scheme the counterterm is then simply 
$S^{(1)}_{\rm ct}=-S^{(1)}_{\rm div}$. If there is a boundary field for 
which $B_{\phi}{}^{p}\ne 0$ and  the resonance condition $y_p=y_\phi$ is 
satisfied, we have a logarithmic divergence which results in 
a universal term linear in $\lambda$ in the boundary beta function \cite{FGK}
\begin{equation}
\beta^{p} = y_{p}\, \mu^{p}+ \lambda\frac{B_{\phi}{}^{p}}{2} + \dots \ .
\end{equation} 
In the case when the bulk perturbation
is marginal the resonance condition requires that the boundary field
is also marginal.

We shall, in the following, always assume that the resonance 
condition is not satisfied, {\it i.e.}\ that $y_p\neq y_\phi$; this is for 
example true in the context of section \ref{sec2.0} where 
$y_{\phi}=0$ and $y_p>0$. Then the counterterm is given by
$S^{(1)}_{\rm ct}=-S^{(1)}_{\rm div}$.
\smallskip

\noindent At the next order in perturbation theory we encounter the integral
\begin{equation}\label{psum}
\sum_{q}\lambda \, \mu^{q}\, l^{-y_{q} - y_{\phi}}\,
\int\!\! dx'\Bigl[  \iint\!\! dxdy  \,  \theta(y-\epsilon/2) \,
\theta(R^{2}-(x-x')^{2}-y^{2}) \, \phi(z,\bar z)\, \psi_{q}(x') \Bigr] \ ,
\end{equation}
where $z=x+iy$ and  $R$ is an infrared regulator. 
The quantity in the square brackets in (\ref{psum}) can be expanded 
in local  boundary fields as in (\ref{multOPE}) and (\ref{exprO}); the coefficients
of these fields can be expressed in terms of certain integrals (see below). 
By the same arguments as above, only coefficients of (primary) relevant 
fields can be divergent as we send $\epsilon \to 0$.

\noindent More precisely, the coefficient with which the primary field $\psi_{p}$  
will appear in (\ref{psum}) equals
\begin{equation} \label{Ipq}
I^{p}_{q}=\iint\!\! dxdy\, \theta(y-\epsilon/2) \,
\theta(R^{2}-(x-x')^{2}-y^{2}) \, \langle \phi(z,\bar z)\psi_{q}(0) \psi_{p}(\infty)\rangle   \ .
\end{equation} 
Using the M\"obius symmetry, the correlation function appearing in this formula 
can be written as
\begin{equation}\label{4ptY}
\langle \phi(z,\bar z)\psi_{q}(0) \psi_{p}(\infty)\rangle=
|z|^{y_{q}-y_p+ y_{\phi}-2} \, \eta^{\delta+y_{\phi}-2}\, (1-\eta)^{(2-y_{\phi}-\delta)/2}\, Y(\eta) \ ,
\end{equation}
where
\begin{equation}
\eta = 1 - \frac{\bar z}{z} \ , \qquad \delta = \frac{1}{3}(4-y_p - y_q - y_{\phi}) 
= \frac{1}{3} (h_p+h_q+\Delta) \ , 
\end{equation}
and
\begin{equation}\label{Y}
Y(\eta)=\left\{
\begin{array}{lr}
\sum_A B_{\phi}{}^{A} D_{qA}{}^{p}
e^{i\frac{\pi}{2}(y_A - y_{\phi}+1 )}F_{\phi \bar{\phi}qp}^A(\eta)&(\re z>0)\ , \\[7pt]
\sum_A B_{\phi}{}^{A} D_{Aq}{}^{p} 
e^{i\frac{\pi}{2}(y_{\phi} - y_A -1 )}F_{\phi\bar{\phi}qp}^A(\eta)&(\re z<0)\ .
\end{array}
\right.
\end{equation}
Here the index $A$ runs over all conformal primaries whose conformal
families appear in the intermediate channel. 
The conformal blocks $F_{\phi\bar{\phi}qp}^A(\eta)$ have a branch cut 
along the real $\eta$-axis from $-\infty$ to $1$ and are normalised so that 
$F^{A}(\eta) \sim \eta^{h_A -\delta}$ with coefficient $1$ as $\eta \to 0$. 
The conformal blocks entering the function $Y(\eta)$ are defined on 
opposite sides of the branch cut for $\re z>0$ and $\re z<0$. The analyticity 
in $z$ implies that the values of $Y(\eta)$ in the lower half $\eta$-plane are 
obtained by the analytic continuation in a clockwise direction 
from the upper half-plane \cite{Lewellen}.    
\smallskip

\noindent Passing to polar coordinates $z=re^{i\vartheta}$ and using (\ref{4ptY}) 
we can rewrite (\ref{Ipq}) as 
\begin{equation}
I_{q}^{p}=\int\limits_{\vartheta_{*}}^{\pi-\vartheta_{*}}\!\!d\vartheta \int\limits_{r_{*}(\eta)}^{R}\!\! dr\, 
 r^{y_{q}-y_p+y_{\phi} - 1}\eta^{\delta+y_{\phi}-2}
(1-\eta)^{(2-y_{\phi}-\delta)/2}Y(\eta) \ ,
\end{equation}
where
\begin{equation}
\vartheta_{*}=\arcsin\left(\frac{\epsilon}{2R}\right)\ , \qquad 
r_{*}^{2}(\eta)=\frac{\epsilon^{2}(\eta-1)}{\eta^{2}} \ .
\end{equation}
Since $\eta=1-e^{-2i\vartheta}$ depends only on $\vartheta$ we can 
perform the integral over $r$. In the remaining integral it is 
convenient to change the  integration variable  $\vartheta$ to $\eta$. 
Altogether we then obtain
\begin{equation}\label{Ieta}
I_{q}^{p}=\frac{i}{2\, \zeta_{pq}}\, \int\limits_{C(\epsilon/R)}\!\!d\eta \Bigl[ 
\epsilon^{\zeta_{pq}}\left(\frac{\eta-1}{\eta^{2}} \right)^{\zeta_{pq}/2} - R^{\zeta_{pq}}  \Bigr]
\eta^{\delta+y_{\phi}-2}(1-\eta)^{-(y_{\phi}+\delta)/2}Y(\eta) \ ,
\end{equation}
where 
\begin{equation}
\zeta_{pq}=y_q-y_p + y_{\phi} \ ,
\end{equation}
and the contour of $\eta$-integration  is a segment of the  circle of radius 1 centered 
at $\eta=1$ and oriented clockwise
\begin{equation}
C(\epsilon/R)=\{\eta=1-e^{-2i\vartheta},\,\, \vartheta_{*}\le \vartheta\le \pi-\vartheta_{*}  \}\ . 
\end{equation}
Obviously, this expression only makes sense if $\zeta_{pq}\neq 0$. In
the resonance case ({\it i.e.}\ for $\zeta_{pq}=0$) we have instead
\begin{equation}\label{Ietar}
(I_{q}^{p})_{\rm res}=\frac{i}{2}\int\limits_{C(\epsilon/R)}\!\!d\eta\, 
\ln\left(\frac{\epsilon}{R|\eta|}\right) 
\eta^{\delta+y_{\phi}-2}(1-\eta)^{-(y_{\phi}+\delta)/2}Y(\eta)\ . 
\end{equation}
\smallskip

The divergences of $I_{q}^{p}$ and $(I_{q}^{p})_{\rm res}$
in the limit $\epsilon\to 0$ can now be analysed 
using well-known properties of conformal blocks. There are two kinds of divergences
that will be important to us: those that come from the region of 
integration $\eta \sim \epsilon/R \to 0$ where the bulk operator approaches 
the boundary far away from the point of insertion of $\psi_{q}$; and those that
arise when the bulk field approaches the boundary in the vicinity of the boundary field 
$\psi_{q}$. As we have argued before (and as will become clear below) the
former divergences are canceled by the contribution from the lower order counterterm
$S_{\rm ct}^{(1)}$, while the remaining divergences have the power 
$\epsilon^{\zeta_{pq}}$. To see this, we use the asymptotics --- see (\ref{Y})
\begin{eqnarray}\label{Yas}
&\mbox{for } \vartheta \to 0 \quad  &Y(\eta)\sim \sum_A B_{\phi}{}^{A} D_{qA}{}^{p}
e^{i\frac{\pi}{2}(y_A - y_{\phi} +1)}\eta^{1-y_A-\delta} + \dots \, , \nonumber \\
&\mbox{for } \vartheta \to \pi \quad  &Y(\eta)\sim 
\sum_A B_{\phi}{}^{A} D_{Aq}{}^{p} 
e^{i\frac{\pi}{2}(y_{\phi} - y_A -1)} \eta^{1-y_A-\delta}
 + \dots
\end{eqnarray} 
in (\ref{Ieta}) and (\ref{Ietar}), and then perform the $\eta$ integrals in the vicinity 
of $\eta=0$,  {\it i.e.}\ from $\eta = i \tfrac{\epsilon}{R}$ up to some intermediate
cut-off $\xi$. This leads to 
\begin{equation}
I_{q}^{p} = C_{q}^{p}\epsilon^{\zeta_{pq}} + f_{q}^{p} R^{\zeta_{pq}}
-\sum_{A}\Bigl[
\frac{B_{\phi}{}^{A}(D_{qA}{}^{p}+D_{Aq}{}^{p})}{2(y_{\phi} -y_A)(y_A + y_q - y_p)}
\epsilon^{y_{\phi} -y_A}R^{y_A+y_q-y_p}   
+ \mathcal{O}_{}(\epsilon^{y_{\phi} - y_A + 1}) \Bigr] \ , \label{Ias} 
\end{equation}
\begin{equation}
(I_{q}^{p})_{\rm res} =  (C_{q}^{p})_{\rm res}\ln(\epsilon/l)  + (f_{q}^{p})_{\rm res}
+\sum_{A}\Bigl[
\frac{B_{\phi}{}^{A}(D_{qA}{}^{p}+D_{Aq}{}^{p})}{2(y_{\phi}-y_A)^{2}}
\left(\frac{\epsilon}{R} \right)^{y_{\phi}-y_A}
+ \mathcal{O}_{}(\epsilon^{y_{\phi} - y_A + 1})\Bigr]  \, , \label{las1}
\end{equation}
where $C_{q}^{p}$, $(C_{q}^{p})_{\rm res}$, $f_{q}^{p}$  and $(f_{q}^{p})_{\rm res}$ 
are some constants independent of $\epsilon$ and $R$ (that come from the evaluation 
of the primitive function at $\xi$, as well as from the remaining part of the integral).  
Since we are 
only interested in divergent contributions in $\epsilon$, we 
may restrict the fields $A$ to be relevant primary fields in
$A \in I^{(1)}$. Furthermore, we can ignore all the subleading terms
$\mathcal{O}_{}(\epsilon^{y_{\phi} - y_A + 1})$ since 
they vanish in the limit $\epsilon\to 0$.\footnote{This is obvious
for $y_A<1$. In a unitary BCFT $y_A=1$ corresponds always to the identity 
operator $\Omega$ which does not have a descendant operator at level 
one since $L_{-1}\Omega=0$.} 
\smallskip

Now we want to show that the divergent terms in the sum in (\ref{Ias}) are precisely
canceled by the lower order counterterm $S_{\rm ct}^{(1)}$ given in (\ref{S1}). At 
order $\lambda \mu^q$ --- recall that $\mu^q$ is 
the coupling constant corresponding to  $\psi_q$ --- the counterterm leads to the
contribution
 \begin{equation}
 \lambda \mu^q\, l^{-y_q-y_{\phi}} \sum_{s\in I^{(1)}}   \epsilon^{y_{\phi}-y_s} \,
 \frac{B_{\phi}{}^{s}}{2(y_{\phi}-y_s)} 
 \iint\!\! dxdx'\, \theta(|x-x'|-\epsilon)\theta(R-|x-x'|) \psi_{s}(x)\psi_{q}(x')\, .   
 \end{equation} 
Again this can be expanded in terms of local boundary fields, and the
divergence in the coefficient of $\psi_p$ equals
\begin{equation}\label{Ict1}
(I_{ct}^{(1)})_{q}^{p} = \sum_{s\in I^{(1)}}
\frac{B_{\phi}^{s}(D_{qs}{}^{p}+D_{sq}{}^{p})}{2(y_{\phi} -y_s)(y_s + y_q - y_p)}
[\epsilon^{y_{\phi} -y_s}R^{y_s+y_q-y_p} - \epsilon^{\zeta_{pq}}]\, . 
\end{equation}
Here we have assumed that there is no resonance among the
boundary fields,  {\it i.e.}\ that $y_s + y_q \ne y_p$ for any $s\in I^{(1)}$;
if there is a resonance, {\it i.e.} $y_s + y_q = y_p$, then (\ref{Ict1}) has 
to be modified in the obvious manner.

In either case, by comparison with (\ref{Ias}), it is now 
clear that the contribution $(I_{ct}^{(1)})_{q}^{p}$ cancels precisely the divergent
part of the sum in (\ref{Ias}), and similarly for the resonant case $\zeta_{pq}=0$. 
Thus the divergent contribution only comes from the first term in (\ref{Ias}) and 
(\ref{las1})
\begin{equation}
\begin{array}{rcl}
\tilde I_{q}^{p} & = & I_{q}^{p} + (I_{ct}^{(1)})_{q}^{p} \  \sim \ 
\tilde C_{q}^{p}\epsilon^{\zeta_{pq}}\, , 
\\
(\tilde I_{q}^{p})_{\rm res}  & = & 
(I_{q}^{p})_{\rm res} + ((I_{ct}^{(1)})_{q}^{p} )_{\rm res} 
\ \sim \  (\tilde C_{q}^{p})_{\rm res}\ln(\epsilon/l) 
\end{array}
\qquad \mbox{as  $\epsilon \to 0$}\ , 
\end{equation} 
where $\tilde C_{q}^{p}$ and $(\tilde C_{q}^{p})_{\rm res}$ are coefficients that can be 
obtained by taking the limits
\begin{equation}
\tilde C_{q}^{p} = \lim_{\epsilon \to 0} \epsilon^{-\zeta_{pq}} \,
\epsilon \partial_{\epsilon } \, 
{\tilde I}_{q}^{p}\, , 
\qquad 
(\tilde C_{q}^{p})_{\rm res} = \lim_{\epsilon \to 0} \epsilon \partial_{\epsilon } 
(\tilde I_{q}^{p})_{\rm res}  \, . 
\end{equation}
Using the explicit expressions (\ref{Ieta}), (\ref{Ietar}) and (\ref{Ict1}) we finally obtain
\begin{eqnarray}\label{C_pq}
 \tilde C_{q}^{p} & =  &
\lim_{\epsilon \to 0} \, \Bigl[ \frac{i}{2\zeta_{pq}}\!
\int\limits_{C(\epsilon/R)}\!\!d\eta\, 
(1-\eta)^{-(y_{\phi} + \delta)/2}(\eta-1)^{\zeta_{pq}/2}
\eta^{y_{p} -y_{q}+ \delta -2}Y(\eta) \\
&& + \sum_{s\in I^{(1)}}
\frac{B_{\phi}{}^{s}(D_{qs}{}^{p}+D_{sq}{}^{p})}{2(y_s + y_q-y_p )\zeta_{pq}}
\left( \frac{R}{\epsilon}\right)^{y_s+y_q-y_p} \Bigr]    + \sum_{s\in I^{(1)}}
\frac{B_{\phi}{}^{s}(D_{qs}{}^{p}+D_{sq}{}^{p})}{2(y_s-y_{\phi} )(y_s + y_q - y_p)} \ ,  
\nonumber  \\[4pt]
( \tilde C_{q}^{p})_{\rm res} & = &
\lim_{\epsilon \to 0} \, \Bigl[ \frac{i}{2}\!
\int\limits_{C(\epsilon/R)}\!\!d\eta\,  
\eta^{y_{\phi} + \delta -2}(1-\eta)^{-(y_{\phi} + \delta)/2}Y(\eta) \nonumber  \\[-16pt]
& & \qquad \qquad \qquad \qquad \qquad\qquad
 + \sum_{s\in I^{(1)}}
\frac{B_{\phi}{}^{s}(D_{qs}{}^{p}+D_{sq}{}^{p})}{2(y_s-y_{\phi} )}
\left( \frac{R}{\epsilon}\right)^{y_s-y_{\phi}} 
\Bigr]\  . \label{Cres}
 \end{eqnarray}
It is worth noting that the contours of the $\eta$-integrations in 
(\ref{C_pq}) and (\ref{Cres}) can be deformed provided the ends of the contour are 
held fixed and the cut is not crossed. In particular one can 
deform $C(\epsilon/R)$ to run infinitesimally above the cut to 
$\eta=1$, and then infinitesimally below back to the second endpoint near $\eta=0$.  
\smallskip

Given that the conformal blocks  only have singularities at $\eta=1,0,\infty$ with 
standard asymptotics, the quantities $\tilde C_{q}^{p}$, $( \tilde C_{q}^{p})_{\rm res}$ 
defined in (\ref{C_pq}) and (\ref{Cres}) are finite and independent of $R$. 
This essentially provides a proof that at order 
$\lambda \mu^p$ the divergences can be canceled by local 
counterterms which are linear combinations of  relevant operators. 
Together with the well known results at orders $\lambda^{2}$ and 
$\mu^{p} \mu^{q}$ we have thus shown that a generic perturbation by 
relevant or marginal bulk and boundary fields is renormalisable at the 
quadratic order in the couplings.\footnote{It should be clear from our 
analysis that the different technical assumptions, namely that 
$y_{s}\ne y_{\phi}$ for any boundary field $\psi_{s}$, and that we only have
a single bulk field, are not crucial for the argument.} It should also
be possible to extend the analysis to higher orders in perturbation theory,
but we have not attempted to do so.

If the resonance condition $\zeta_{pq}=0$ is not satisfied the 
divergence of $\tilde I_{q}^{p}$ is power like, and in the minimal 
subtraction scheme there are no terms of order $\lambda \mu^p$ in 
the beta function $\beta^p$. On the other hand, when the resonance 
condition is satisfied for a pair $(p,q)$ we have a universal term 
({\it cf.}\ (\ref{ex1}) and (\ref{ex2}))
\begin{equation}
\beta^{p}=y_{p}\, \mu^{p} - ( \tilde C_{q}^{p})_{\rm res}\, \lambda \, \mu^{q} + \dots  \ .
\end{equation}
For a  marginal bulk perturbation $y_{\phi}=0$, the $p=q$ case (and only that one) 
is always resonant so that we have in the notation of section \ref{sec2.0} 
\begin{equation}
{\mathcal{E}}_{\phi q}^{p}= ({\mathcal{E}}_{\phi q}^{p})_{\rm min}\equiv\left\{ \begin{array}{ll}
-( \tilde C_{p}^{p})_{\rm res} & \mbox{for $p= q$} \\
0 & \mbox{for $p\ne q$.}  
\end{array} \right.
\end{equation}
Moreover, assuming as in section \ref{sec2.0} that all $y_{p}>0$, we have in the 
minimal subtraction scheme 
\begin{equation}
\mathcal{D}_{pr}^{p}=\mathcal{D}_{rp}^{p}=0\, , \quad \mathcal{B}_{\phi}^{p}=0  \ ,
\end{equation}
and therefore by (\ref{dim_shift}) 
\begin{equation}\label{EC}
\tilde {\mathcal{E}}_{\phi p}^{p} = -( \tilde C_{p}^{p})_{\rm res} \ ,
\end{equation}
where $( \tilde C_{p}^{p})_{\rm res}$ is given by formula (\ref{Cres}) for $y_{\phi}=0$ 
and $p=q$.   As  we proved in section \ref{sec2.0}, the quantity 
$\tilde {\mathcal{E}}_{q}^{p}$ is scheme independent;  we have therefore
managed to obtain a description of this universal quantity in terms of 
conformal blocks --- this is the main result of this subsection.

\subsection{Computation in a Wilsonian scheme}

It is instructive to compute  (\ref{dim_shift}) also in a different, Wilsonian type, 
renormalisation scheme --- the one of \cite{Cardy,AL, FGK}. 
We will refer to this scheme as the `OPE scheme' for the reason that the 
first nontrivial terms in the beta functions are given by various OPE coefficients. 
This scheme is often employed in conformal perturbation theory at the leading 
order. One of the advantages of this scheme is that in the presence of a nearby 
infrared fixed point in theory space, the corresponding coordinates are nonsingular 
near that fixed point. Another attractive feature is that formulae for  universal quantities, 
such as the dimension shift (\ref{dim_shift}), can be obtained quite easily in contrast 
with the minimal subtraction scheme. On the other hand we will see at the end of this 
section that the scheme has some pitfalls when applied to computing non-universal 
quantities at higher order in perturbation theory.

In the OPE  scheme the theory is also regulated by a point-splitting 
cut-off $\epsilon$,  just as in the minimal subtraction scheme of the previous section. 
It is however convenient to introduce the infrared regulator slightly differently:
we introduce a cut-off  whenever two coordinates $x_{i}$, $x_{j}$ are separated 
by a distance larger than $L$, and whenever there is a bulk operator at 
a distance $y>L$.  
The dimensionless couplings are now introduced by using the UV cut-off 
scale itself, which is  understood as a fundamental UV scale (lattice spacing, 
atomic or molecular scale). Thus we have
\begin{equation}\label{deltaS2}
\delta S=\sum_k \epsilon^{\Delta_k-2}\lambda^k\iint\!\! dxdy\,\phi_k(x,y)
+\sum_{p}\epsilon^{h_p-1}\mu^{p} \int\!\! dx\,\psi_p(x)\ .
\end{equation}
Note that one can also include in (\ref{deltaS2}) irrelevant 
operators $\psi_{A}$ with $y_{A}<0$. Their 
contributions will be relatively suppressed as $\epsilon^{-y_{A}}$ 
but one may worry that in the perturbation expansion they will lead to 
contributions more singular than this suppression factor. We will see 
that, although there is no need to introduce irrelevant operators at the 
leading order, they are sometimes necessary to be taken into account 
at higher orders in perturbation theory.

We will confine ourselves throughout this subsection to the case of a single 
marginal bulk field $\phi(x,y)$ ($\Delta=2$).  In this case the terms in the 
perturbation expansion we are interested in are
\begin{eqnarray}
 e^{\delta S}=&& 1 + \lambda \iint \!\!dxdy\, \phi(x,y)\theta(y-\frac{\epsilon}{2}) 
+ \sum_p \epsilon^{-y_p}\mu^p \int\!\! dx\, \psi_{p}(x) 
\nonumber \\
&& + \sum_p \epsilon^{-y_p}\mu^p \lambda 
\iint\!\! dxdy \theta(y-\frac{\epsilon}{2})\int\!\! dx'\, \phi(x,y)\psi_{p}(x') \,
\theta(L-|x-x'|)\, \theta(L-y)
 \\
&& + \sum_{pq}\epsilon^{-y_p-y_q}\mu^p \mu^q  \iint\!\! dx_1 dx_2 
\psi_p(x_1)\psi_q(x_2)\, \theta(x_2-x_1 -\epsilon) \, 
 \theta(L-|x_1-x_2|) + \dots \ .
\nonumber
\end{eqnarray}
The cut-off variation 
$\epsilon \partial_\epsilon e^{\delta S}$ can be computed assuming that the coupling 
constants depend on the cut-off via the couplings themselves according to 
\begin{equation}
\epsilon \partial_\epsilon \mu^p = \beta^p(\mu^q, \lambda) \ ,
\end{equation}
where $\beta^{p}$ are the beta functions (\ref{genbdrybeta}). In the OPE
scheme we now vary $e^{\delta S}$ with respect to $\epsilon$, {\it i.e.}\
we compute $\epsilon \partial_\epsilon e^{\delta S}$, and demand
that the variation vanishes at the leading order in $\epsilon$. This reflects
the main principle of the Wilsonian renormalisation group approach,
namely that the renormalised quantities must be independent of the UV scale.
The resulting equations fix order by order the coefficients of the beta functions.  
The linear terms in the beta functions are always scheme independent with 
the coefficients given by the anomalous dimensions.
It is easy to check that at the linear order in $\mu^p$ the equation 
\begin{equation}\label{scaleindp}
\epsilon \partial_\epsilon e^{\delta S} \esim 0 
\end{equation}
is satisfied automatically. The equation arising at  the linear order in $\lambda$  
fixes 
\begin{equation}\label{calB}
{\mathcal B}_{\phi}^{ p}=\frac{1}{2}B_{\phi}{}^{p} \ ,
\end{equation}
where $B_{\phi}{}^{p}$ are the bulk-to-boundary OPE coefficients 
(\ref{bulkboundaryOPE}) (see \cite{FGK}). At the quadratic order in the boundary 
couplings  one obtains the well known expression
\begin{equation}\label{calD}
{\mathcal D}_{rs}^{p}= D_{rs}{}^{p} \ ,
\end{equation} 
where $D_{rs}{}^{p}$ are the boundary OPE coefficients (\ref{boundaryboundaryOPE}).
Finally, the equation at order $\lambda\mu^q$ is
\begin{eqnarray}\label{lambdamu}
0&\esim & \lambda\mu^q \Bigl[ - \frac{\epsilon^{1-y_q}}{2}
\iint\!\! dxdx'\, \phi(x,\frac{\epsilon}{2})\psi_q(x') \,
\theta(L-|x-x'|) \nonumber \\
 && + \sum_{p} \frac{\epsilon^{-y_p-y_q}}{2} B_{\phi}{}^{p}
\iint\!\!dx_1 dx_2 \, 
\psi_q(x_1)\psi_{p}(x_2) \, 
\theta(|x_1-x_2|-\epsilon) \theta(L-|x_1-x_2|) \nonumber \\
&& + \sum_{p} \epsilon^{-y_p}{\mathcal{E}}_{\phi q}^{p} \int\!\!dx\, \psi^{p}(x)  \Bigr]\ .
\end{eqnarray}
 The first line in the above expression came from applying $\epsilon\partial_\epsilon$ to the cut-off function $\theta(y-\frac{\epsilon}{2})$ while the
second line  came from the lower order term (\ref{calB}). 
The coefficients ${\mathcal{E}}_{\phi q}^{p}$ 
are formally obtained by taking correlation functions with the operator   $\psi_{p}$ inserted at 
infinity
\begin{eqnarray} \label{exp4pt}
{\mathcal{E}}_{\phi q}^{p}= ({\mathcal{E}}_{\phi q}^{p})_{\rm OPE}&\equiv  &
\frac{1}{2}\lim_{\epsilon \to 0} \epsilon^{y_{p}-y_q} 
\Biggl[ \epsilon \int\limits_{-L}^{L}\!\! dx\,  \langle \phi(x,\frac{\epsilon}{2})
\psi_{q}(0) \psi_{p}(\infty)\rangle \nonumber \\ && 
\quad - \sum_{r}\epsilon^{-y_{r}} B_{\phi}{}^{r}
\int\limits_{-L}^{L}\!\!dx\,  \langle 
\psi_{q}(0) \psi_{r}(x)  \psi_{p}(\infty)
\rangle \, \theta (|x| - \epsilon)
\Biggr]\ .
\end{eqnarray}
The above expression is formal because the limit may not exist. 
The integrals in the second line in (\ref{exp4pt}) can be evaluated explicitly
\begin{eqnarray}\label{expl_int}
&&\frac{1}{2}\sum_{r}\epsilon^{y_p-y_q-y_{r}} B_{\phi}{}^{r}
\int\limits_{-L}^{L}\!\!dx\,  \langle 
\psi_{q}(0) \psi_{r}(x) \psi_{p}(\infty)
\rangle\,  \theta(|x|-\epsilon)\nonumber \\
&&\qquad\qquad = 
\sum_{r}B_{\phi}{}^{r}\frac{D_{(qr)}{}^{p}}{(y_q-y_p+y_{r})}\, 
\left[\left(\frac{L}{\epsilon}\right)^{y_q-y_p+y_r} -1\right] \equiv H(\epsilon/L)
\ . 
\end{eqnarray}
To study the convergence we 
rewrite  expression  (\ref{exp4pt}) 
 via conformal blocks using (\ref{4ptY})
 \begin{equation}\label{expY2}
 ({\mathcal{E}}_{\phi q}^{p})_{\rm OPE}=\lim_{\epsilon \to 0}\Bigl[ \frac{i}{2}
 \int\limits_{C'(\epsilon/L)}\!\!d\eta\, \eta^{\delta-y_q+y_p-2}
 (1-\eta)^{-\delta/2}(\eta-1)^{(y_p-y_q)/2}
 Y(\eta)  - H(\epsilon/L)
 \Bigr] \ ,
 \end{equation}
 where 
 \begin{equation}
C'(\epsilon/R)=\{\eta=1-e^{-2i\vartheta}, \vartheta_{*}'\le \vartheta\le \pi-\vartheta_{*}'  \}\, , 
\quad  \vartheta_{*}'=\frac{1}{2}\ln\left(\frac{1-i\epsilon/2L}{1+i\epsilon/2L}  \right)
\end{equation}
is a segment of a unit circle around $\eta=1$ oriented clockwise. If we now
evaluate $\tilde {\mathcal{E}}_{\phi p}^{p}$ of (\ref{dim_shift}) in the OPE
scheme, using (\ref{expY2}) for $p=q$, as well as 
(\ref{calD}) and (\ref{calB}), we obtain the same expression 
in terms of conformal blocks as given in 
(\ref{Cres}) and  (\ref{EC}). Thus  $\tilde {\mathcal{E}}_{\phi p}^{p}$ is indeed
scheme-independent, as we have argued before. In terms of correlation functions,
it can now be written as 
\begin{equation}\label{Wilson_E}
\tilde {\mathcal{E}}_{\phi p}^{p} = 
\lim_{\epsilon \to 0}  
\left[ \frac{\epsilon}{2} \int\limits_{-L}^{L}\!\! dx\,  \langle \phi(x,\frac{\epsilon}{2})
 \psi_{p}(0) \psi_{p}(\infty)\rangle
 -\sum_{r} D_{(pr)}{}^{p} \frac{B_{\phi}{}^{r}}{y_{r}}
\left(\frac{L}{\epsilon}\right)^{y_r} \right] \ .
 \end{equation}
Writing $\delta=\tfrac{\epsilon}{2L}$ and using the variable $\eta=1-e^{-2i\theta}$ in
the integral (\ref{expY2}), we can also obtain another, perhaps more elegant 
expression for $\tilde {\mathcal{E}}_{\phi p}^{p}$ 
\begin{equation}\label{WilsonE2}
\tilde {\mathcal{E}}_{\phi p}^{p} = 
\lim_{\delta \to 0}  
\left[  \int\limits_{ \delta}^{\pi-\delta}\!\! d\vartheta\,  
\langle \phi(e^{i\vartheta})
\psi_{p}(0) \psi_{p}(\infty)\rangle
-\sum_{r} D_{(pr)}{}^{p} \frac{B_{\phi}{}^{r}}{y_{r}}
\left(\frac{1}{2\delta}\right)^{y_r} \right] \ ,
\end{equation}
where   the bulk field insertion runs 
over a semicircle of radius one around the boundary insertion $\psi_{p}(0)$. 
Note that there is nothing special about  the radius being one, since
\begin{equation}
\langle \phi(e^{i\vartheta})
\psi_{p}(0) \psi_{p}(\infty)\rangle = 
\rho^2 \, \langle \phi(\rho e^{i\vartheta})
\psi_{p}(0) \psi_{p}(\infty)\rangle
\end{equation}
for any $\rho>0$. 
\smallskip

Let us now come back to the expression (\ref{expY2}) for $p\ne q$. Substituting 
the asymptotic expansion
(\ref{Yas}) into (\ref{expY2}) we find that although the most dangerous 
divergences,  associated with the leading contributions of relevant primaries in (\ref{Yas}), 
cancel out, there may be divergences coming from terms in (\ref{Yas}) associated with 
irrelevant fields. More precisely there are additional divergences in 
$({\mathcal{E}}_{\phi q}^{p})_{\rm OPE}$ from the region near $\eta \sim  0$ whenever 
there is an irrelevant primary $\phi_{A}(z)$ in the theory such that 
\begin{equation}\label{1div}
B_{\phi}{}^{A}\ne 0 \quad \mbox{and} \quad \{ D_{qA}{}^{p}\ne 0  \mbox{ or }  
D_{Aq}{}^{p}\ne 0 \} \quad \mbox{and} \quad y_{A} + y_{q}> y_{p} \ ,
\end{equation}
or whenever we have a relevant primary $\psi_{r}(z)$ such that 
\begin{equation}\label{2div}
B_{\phi}{}^{r}\ne 0 \quad \mbox{and} \quad \{ D_{qr}{}^{p}\ne 0   \mbox{ or }  
D_{rq}{}^{p}\ne 0 \} \quad \mbox{and} \quad  y_{r} + y_{q}> y_{p} + 1 \ .
\end{equation}
In the last case the irrelevant field  causing the divergence is a descendant 
of the primary $\psi_{r}$. From the point of view of the minimal subtraction scheme
of the previous subsection, this problem does not arise since 
the conditions (\ref{1div}) and (\ref{2div}) imply that there are no 
divergences from the region where the bulk field approaches the boundary 
insertion.
In fact, the additional $\epsilon \to 0$ divergences in the OPE scheme come 
directly from the extra divergent factors of $\epsilon$ included in the action 
(\ref{deltaS2}). 

The situation can be mended if we include in the original perturbed action
(\ref{deltaS2}) also irrelevant fields, and introduce their beta functions by 
requiring that $\epsilon \partial_{\epsilon} e^{\delta S} \sim 0$  at the subleading 
orders in $\epsilon$. Then the function $H(\epsilon/L)$ is modified accordingly to 
include more divergent terms that cancel out the divergences 
coming from the integral in (\ref{expY2}). 
Although this resolution looks quite natural from the Wilsonian point of view, the whole 
scheme becomes quite unwieldy for practical applications whenever 
(\ref{1div}) or (\ref{2div}) happens.  Note, however, that these 
extra divergences do {\em not} appear for universal quantities like 
$\tilde {\mathcal{E}}_{\phi p}^{p}$. Thus, as long as we are only interested in 
these quantities we can (and will) use the technically simpler OPE scheme. 
In particular, we will use this method to compute analogous quantities for 
pure bulk and pure boundary perturbations in section~4.

It is worth noting that  the complications related to (\ref{1div}) and (\ref{2div})  arise 
only in the presence of  several running coupling constants. Although beta function 
coefficients were studied for some models to a very large order, see {\it e.g.}\ 
\cite{Ludwig:2002fu}, such computations  
typically involved  only a single coupling constant. 
 

\subsection{Perturbations by boundary changing operators}
 
Up to now we have assumed that there is a single (fundamental) 
boundary condition. The whole analysis can be easily generalised to 
the situation where we have superpositions of boundary conditions;
in that case the set of boundary operators includes also boundary
changing operators $\psi_{p}^{ab}(x)$, where the
two boundary conditions are labeled by $a$ and $b$ with 
$a$ being the boundary condition to the left of $x$ and $b$ to the right. 
Local excitations of the pure boundary $a$ are denoted $\psi_p^{aa}$.
The study of renormalisation group flows involving such operators was initiated in 
\cite{Graham}.

The OPEs of  a bulk field approaching the boundary with
label $a$, and that of two boundary fields have the form
\begin{eqnarray}
\phi_i(x+iy,x-iy)&=&\sum_{r}\bi{B}{i}{r}{a\, }
(2y)^{h_r-\Delta_i}\psi_{r}^{aa}+\ldots\,,
\label{bulkboundaryOPE2}\\[4pt]
\psi_p^{ab}(x)\psi_q^{bc}(y)&=&\sum_{r}D_{pq}^{(abc)r}
(y-x)^{h_r-h_p-h_q}\psi_r(y)+\ldots \qquad (y>x) \ . 
\label{boundaryboundaryOPE2}
\end{eqnarray} 
The only difference to the previous analysis is that there are now
various superselection rules that demand for example, that
products of boundary operators can only be non-zero if the 
intermediate boundary conditions match, or that the boundary
fields that appear in the bulk to boundary OPE are always
boundary preserving fields. Taking this into account, the boundary 
beta functions then have the following general form
\begin{equation}
\beta_p^{ab}=y_p^{ab}\mu^{p(ab)}+{}^{a\!}\mathcal{B}_{\phi}^{p}
\,\delta^{ab}\lambda+ \sum_{c;rs}\mathcal{D}_{rs}^{p(acb)}\mu^{r(ac)}
\mu^{s(cb)}+\sum_{r}\mathcal{E}_{\phi r}^{p(ab)}\lambda\mu^{r(ab)}+\ldots \ ,
\label{genbdrybeta2}
\end{equation}
where $y_p^{ab}=1-h_{p}^{ab}$ are anomalous dimensions and $\mu^{r(ab)}$ 
are the coupling constants of the operators $\psi^{ab}_{r}(x)$.
The expression for the dimension shift (\ref{dim_shift}) generalises as 
\begin{equation}\label{dimshift}
\tilde {\mathcal{E}}_{\phi p}^{p(ab)}={\mathcal{E}}_{\phi p}^{p(ab)} - 
\sum_{r} \mathcal{D}_{rp}^{p(aab)}
\frac{ {}^{a\!}\mathcal{B}_{\phi}^{r}}{y_{r}^{aa}} - 
\sum_{r} \mathcal{D}_{pr}^{p(abb)}
\frac{{}^{b\!}\mathcal{B}_{\phi}^{r} }{y_{r}^{bb}}\  .
\end{equation}
Similarly, the main results of the previous subsection 
(\ref{Wilson_E}) and (\ref{WilsonE2}) now become
\begin{eqnarray}
\tilde {\mathcal{E}}_{\phi p}^{p(ab)} & = &
\lim_{\epsilon \to 0}  
\frac{1}{2}\Biggl[ \epsilon \int\limits_{-L}^{L}\!\! dx\,  \langle \phi(x,\frac{\epsilon}{2})
 \psi_{p}^{ab}(0) \psi_{p}^{ba}(\infty)\rangle \nonumber \\
 && \quad\qquad  -\sum_{r}
 D_{rp}^{(aab)p} \frac{\bi{B}{\phi}{r}{a\, } }{ y_{r}^{aa}}
\left(\frac{L}{\epsilon}\right)^{y_r^{aa}} 
-\sum_{r}D_{pr}^{(abb)p} \frac{\bi{B}{\phi}{r}{b\, } }{ y_{r}^{bb}}
\left(\frac{L}{\epsilon}\right)^{y_r^{bb}}
\Biggr] \ ,  \label{Wilson_E2} \\
\tilde {\mathcal{E}}_{\phi p}^{p(ab)} & = &
\lim_{\delta \to 0}  
\Biggl[  \int\limits_{\delta}^{\pi-\delta}\!\! d\vartheta\,  
\langle \phi(e^{i\vartheta})
 \psi_{p}^{ab}(0) \psi_{p}^{ba}(\infty)\rangle \nonumber \\
 && \qquad \qquad -\sum_{r}
 D_{rp}^{(aab)p} \frac{ \bi{B}{\phi}{r}{a\, }}{2y_{r}^{aa}}
\left(\frac{1}{2\delta}\right)^{y_r^{aa}} 
-\sum_{r}D_{pr}^{(abb)p}
\frac{\bi{B}{\phi}{r}{b\, } }{2y_{r}^{bb}}
\left(\frac{1}{2\delta}\right)^{y_r^{bb}} \label{Wilson_E3}
\Biggr] \ .
 \end{eqnarray}



\section{Some explicit examples}\label{sec_3}
\setcounter{equation}{0}

Up to now our analysis has been very general. In this section we want to illustrate 
these general results with two simple examples. 

\subsection{A single Neumann brane}

The simplest example is the case of a single Neumann brane on a circle of 
radius $R$. The action of this theory is simply\footnote{Throughout this section
we set $\alpha'=1$.}
\begin{equation}
S=\frac{1}{2\pi}\int d^2z\,\partial X\, \bar{\partial} X\ .
\end{equation}
The bulk field that corresponds to changing the radius $R$ is 
$\phi(z,\bar{z})
=2\partial X(z)\bar{\partial}X(\bar{z})$, which is an 
exactly marginal operator in the bulk. More specifically, we 
shall consider the
perturbation 
\begin{equation}\label{radiusc0}
\delta S=2\lambda \int\!\! dxdy\, \partial X(w)\bar{\partial} X(\bar{w}) 
\end{equation}
that changes the radius $R$ as $R^{\lambda}=R e^{-\pi\lambda}$ so that 
to the first order we have $\delta R=-\pi R\lambda$. 

On the Neumann brane we have open string momentum states
corresponding to the vertex operators $\psi=e^{ikX}$, whose conformal dimension
is $h=k^2$ with $k=\tfrac{n}{R}$ and $n\in {\mathbb Z}$. 
We want to study how the conformal dimension of these operators changes
as we change the radius. Thus we need to calculate\footnote{Note that
$\psi$ is not a self-conjugate field, and we therefore have to insert the conjugate
field at infinity.}
\begin{equation}\label{qq1}
{\cal E} =  \lim_{\delta\to 0} \Biggl[
2 \int_{\delta}^{\pi - \delta} d\vartheta 
\langle  e^{-ik X}(\infty) \, 
\partial X(e^{i\vartheta}) \, \bar\partial X(e^{-i\vartheta}) \,
e^{ik X}(0) \, 
\rangle - \sum_r D_{r\psi}{}^{\psi} \frac{B_\phi{}^{r}}{ y_r} 
\left( \frac{1}{2\delta} \right)^{y_r} \Biggr] \ . 
\end{equation}
On the Neumann boundary we have $\partial X = \bar \partial X$, and 
the correlation function equals
\begin{equation}
2 \langle  \,  e^{-ik X}(\infty) \,  \partial X(e^{i\vartheta}) \, \bar\partial X(e^{-i\vartheta}) \,
e^{ik X}(0) \,  \rangle 
= - 2\, k^2 - \frac{1}{(z-\bar{z})^2} \ .
\end{equation}
Thus the integral is simply 
\begin{eqnarray}
2 \int_{\delta}^{\pi - \delta} d\vartheta \, 
\langle \, e^{-ik X}(\infty) \
 \partial X(e^{i\vartheta}) \, \bar\partial X(e^{-i\vartheta}) \,
e^{ik X}(0) \,  \rangle & =  & - 2 \, k^2 \pi 
-  \left. \frac{1}{4} \cot\vartheta \right|_{\vartheta=\delta}^{\vartheta=\pi-\delta} \nonumber  \\
& = & - 2\, k^2 \pi + \frac{1}{2\delta} + {\cal O}(\delta) \ . 
\end{eqnarray}
The term that is singular in $\delta$ is subtracted by the last term in (\ref{qq1}). In
fact, the only relevant or marginal boundary field that is switched on is the 
identity field with $y_0=1$ and $D_{{\bf 1}\psi}{}^{\psi}=1$, and the 
corresponding bulk to boundary OPE coefficient is $B_\phi{}^{{\bf 1}} = 1$ since
\begin{equation}
2 \langle \partial X (z) \, \bar\partial X(\bar{z}) \rangle = - \frac{1}{(z-\bar{z})^2} 
= \frac{1}{4 y^2} \ .
\end{equation}
Thus we find that ${\cal E} = - 2\, \, k^2 \pi$ in this example, which implies that 
\begin{equation}
\delta h = 2\, k^2 \pi \lambda \ .
\end{equation}
This then agrees with the geometrical expectation since for $h= k^2$
with $k=\tfrac{n}{R}$ we have 
\begin{equation}
\delta h = - 2 k^2 \, \frac{\delta R}{R} =  2\, k^2\, \pi \lambda \ .
\end{equation}

\subsection{Branes at angles}

 A somewhat more interesting example is the configuration of two D1-branes that 
stretch diagonally across a $2$-torus, crossing each other at an angle
(see figure~1). This brane configuration is  obviously unstable since the relative 
open string between the two D1-branes is 
tachyonic but this will not be important in the following. 
(One can imagine that this is only part of 
a more complicated background involving additional directions, and that
the boundary conditions of these D-branes in the other directions are chosen 
so that the relative open string is not tachyonic.) 

\vspace{1cm}
\begin{center}
\epsfig{file=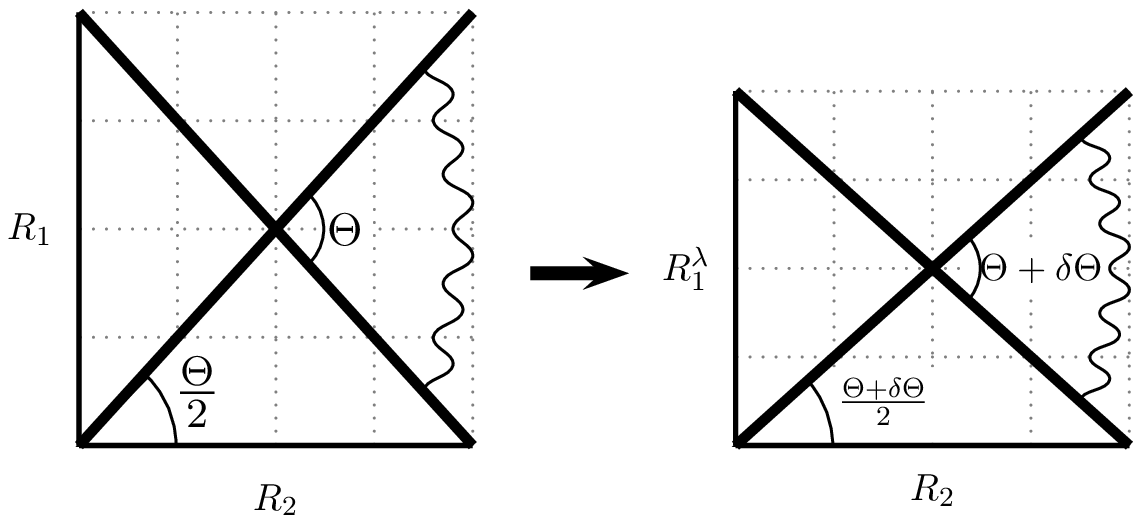, width=10cm}
\end{center}
\textbf{Fig.~1: }\textit{Radius perturbation on the torus: changing
the radius of $R_1$ modifies the relative angle between the two branes,
and hence the conformal dimension of the corresponding
boundary changing field.}
\vspace{1cm}

For simplicity consider the situation where the $T^2$ torus is orthogonal
with radii $R_1$ and $R_2$. The bulk operator that changes either
radius is an exactly marginal bulk operator, but it does have an important
impact on the boundary theory since the ratio of the two radii determines
the conformal dimension of the lowest string excitation between the two D1-branes. 
In the following (section~3.2.2) we shall calculate the change in conformal 
dimension using the RG formalism we have developed above. 
As we shall see, this will reproduce the standard formula for the conformal 
dimension of boundary fields on branes at angles that will be reviewed in 
section~3.2.1.

\subsubsection{The geometrical analysis}

The torus theory is described by the action
\begin{equation}
S=\frac{1}{2\pi}\int d^2z\,\partial X^\mu\bar{\partial} X_\mu\ .
\end{equation}
We shall now consider changing the radius $R_1$ by means of the perturbation
\begin{equation}\label{radiusc}
\delta S=2\lambda \int\!\! dxdy\, \partial X^1(w)\bar{\partial} X^1(\bar{w})  \ ,
\end{equation}
which to first order gives $\delta R_1=-\pi R_1\lambda$. 

The two D-branes stretch diagonally across
the torus; their angle relative to the $x^2$ axis will be denoted by 
$\pm \Theta/2$, where $\Theta$ satisfies
\begin{equation}\label{angle}
\tan\frac{\Theta}{2}=\frac{R_1}{R_2} \ .
\end{equation}
The open string that stretches between the two branes satisfies the 
$+$ boundary condition
\begin{eqnarray}\label{bc+}
\partial X^1(z)&=&-\cos\Theta\,\bar{\partial}X^1(\bar{z})
+\sin\Theta\,\bar{\partial}X^2(\bar{z}) \, , \nonumber \\
\partial X^2(z)&=&\sin\Theta\,\bar{\partial}X^1(\bar{z})
+\cos\Theta\,\bar{\partial}X^2(\bar{z})
\end{eqnarray}
at one end (say for $z=\bar{z}$ on the positive real axis), and the
$-$ boundary condition
\begin{eqnarray}\label{bc-}
\partial X^1(z)&=&-\cos\Theta\,\bar{\partial}X^1(\bar{z})
-\sin\Theta\,\bar{\partial}X^2(\bar{z}) \, , \nonumber \\
\partial X^2(z)&=&-\sin\Theta\,\bar{\partial}X^1(\bar{z})
+\cos\Theta\,\bar{\partial}X^2(\bar{z})
\end{eqnarray}
at the other (say for $z=\bar{z}$ on the negative real axis). By going to complex
variables, {\it i.e.}\ by writing $Z^+=\frac{1}{\sqrt{2}}(X^1+iX^2)$, 
$Z^-=\frac{1}{\sqrt{2}}(X^1-iX^2)$, we can write the open string fields as 
\begin{eqnarray}
&&Z^+(z,\bar{z})=i\sqrt{\frac{1}{2}}\sum_{m\in\mathbb{Z}}
\left(\frac{a^+_{m-\nu}}{(m-\nu)z^{m-\nu}}-e^{-i\Theta}
\frac{a^-_{m+\nu}}{(m+\nu)\bar{z}^{m+\nu}}\right)\ \label{complex}\\
&&Z^-(z,\bar{z})=i\sqrt{\frac{1}{2}}\sum_{m\in\mathbb{Z}}
\left(\frac{a^-_{m+\nu}}{(m+\nu)z^{m+\nu}}-e^{i\Theta}
\frac{a^+_{m-\nu}}{(m-\nu)\bar{z}^{m-\nu}}\right)\ , \label{complex1}
\end{eqnarray}
where $\nu=\Theta/\pi \in [0,1)$. The modes $a^{\pm}_{m\mp\nu}$ satisfy the
canonical commutation relations 
\begin{equation}\label{com}
[a^+_{m-\nu},a^-_{n+\nu}]=(m-\nu)\delta_{m,-n}\ ,
\end{equation}
and the Virasoro generators can be expressed in terms of them as 
\begin{equation}
L_m=\sum_{k\in\mathbb{Z}}:a^+_{m-k-\nu}a^-_{k+\nu}:
\;+\;\tfrac{1}{2}\nu(1-\nu)\delta_{m,0}\ .
\end{equation}
The conformal dimension of the lowest boundary changing operator 
$\psi^{-+}$ is thus 
\begin{equation}\label{Casimirenergy}
h_\psi^{-+}=\frac{1}{2}\nu(1-\nu) \ .
\end{equation}

According to the analysis of \cite{FGK,FGK2}, the two D-branes will respond to
the radius changing bulk perturbation (\ref{radiusc}) by simply adjusting
themselves infinitesimally, so that they continue to stretch diagonally
across. To first order in $\lambda$, the angle $\Theta$ thus
changes via (\ref{angle}) as 
\begin{equation}
\delta\Theta=-\pi\sin\!\Theta\;\lambda \ .
\end{equation}
With $\pi\nu=\Theta$  this implies that the conformal
dimension of the  lowest boundary changing operator
changes as 
\begin{equation}\label{geometric_dh}
\delta h_\psi^{-+}=\frac{1}{2}(2\nu-1)\sin\!\Theta\;\lambda+
\mathcal{O}(\lambda^2)\ .
\end{equation}
This is the result we now want to reproduce using the RG approach
explained above.

\subsubsection{The RG approach}

Formula (\ref{Wilson_E3}) applied to the situation at hand reads
\begin{eqnarray} \label{E4}
\tilde {\mathcal{E}}_{\phi\psi}^{\psi(-+)}&=&\lim_{\delta \to 0}\Bigl[  
2\int\limits_{\delta}^{ \pi-\delta}\!\! d\vartheta\, 
 \langle \partial X^1(e^{i\vartheta})\bar{\partial}X^1(e^{-i\vartheta}) \psi^{-+}(0)\, 
 \psi^{+-}(\infty)\rangle \\
 && \qquad \qquad -\sum_{r}
 D_{r\psi}^{(--+)\psi} \frac{ \bi{B}{\phi}{r}{-\, }}{2y_{r}^{--}}
\left(\frac{1}{2\delta}\right)^{y_r^{--}} 
-\sum_{r}D_{\psi r }^{(-++)\psi}
\frac{\bi{B}{\phi}{r}{+\, } }{2y_{r}^{++}}
\left(\frac{1}{2\delta}\right)^{y_r^{++}}   \nonumber 
\Biggr] \ , 
\end{eqnarray}
where the index $r$ runs over all relevant boundary operators in the 
respective sectors. Since the model at hand is Gaussian   the 
only relevant operator induced on the boundary by $\phi(z,\bar z)$ is the identity operator 
in the respective $+$ or $-$ sector. The corresponding bulk to boundary OPE coefficients can 
be read off from the expectation values 
\begin{equation}\label{Bs}
2\langle \partial X^1(z)\bar{\partial}X^1(\bar z) \rangle_{\pm}=
\frac{\bi{B}{\phi}{\bf 1}{\pm \, } }{4y^{2}}=-\frac{\cos\Theta}{4y^2}\ 
\end{equation} 
which can be computed using the mode expansions  (\ref{complex})
and (\ref{complex1}) for $\nu=0$. (The string fields in the presence of a 
single boundary are of the same form as  (\ref{complex})
and (\ref{complex1}) but with $\nu=0$.) Thus 
\begin{equation}\label{BBs}
\bi{B}{\phi}{\bf 1}{+ \, }=\bi{B}{\phi}{\bf 1}{- \, }=-\cos\Theta  \ .
\end{equation}

The three-point correlator in (\ref{E4}) is given by the one-point function of the radius changing 
operator in the presence of the boundary conditions  (\ref{bc+})
and  (\ref{bc-}). A straightforward computation yields  
\begin{eqnarray}\label{correlator_in_z}
2\langle \partial X^1(z)\bar{\partial}X^1(\bar z)  \psi^{-+}(0)\, 
 \psi^{+-}(\infty)\rangle & = & 
 2\langle \partial X^1(z)\bar{\partial}X^1(\bar{z}) \rangle_{\Theta}  \nonumber \\ 
 &=& \frac{1}{2}e^{-i\Theta}\frac{z^\nu}{\bar{z}^\nu}
\frac{z(1-\nu)+\bar{z}\nu}{z(z-\bar{z})^2}+c.c.
\end{eqnarray}
Viewed as a function on $\mathbb{C}$ rather than on $\mathbb{H}^+$, 
the correlator has a logarithmic branch cut along the negative real 
axis. 
The  integral at hand can be easily evaluated 
\begin{eqnarray}\label{integral}
2\int\limits_{\delta}^{ \pi-\delta}\!\! d\vartheta\,  
 \langle \partial X^1(z)\bar{\partial}X^1(\bar{z}) \rangle_{\Theta} 
 & = &
  - \int\limits_{\delta}^{\pi- \delta}\!\! d\vartheta\, \frac{e^{-i\Theta+2i\nu\vartheta}}{8\sin^2\vartheta}
(1-\nu+\nu e^{-2i\vartheta}) + c.c. \\[10pt]
&= & \displaystyle{\frac{\cos(-\Theta+\vartheta(2\nu-1))}%
{4\sin\vartheta}\bigg|^{\vartheta=\pi-\delta}%
_{\vartheta=\delta}} =
-\frac{\cos(-\Theta+\delta(2\nu-1))}%
{2\sin\delta}  \ ,\nonumber
\end{eqnarray}
where in the last step we  used $\Theta=\pi\nu$.
Substituting (\ref{integral}) and (\ref{BBs}) into  (\ref{E4}) we obtain 
\begin{equation}
\tilde {\mathcal{E}}_{\phi\psi}^{\psi(-+)} = \lim_{\delta \to 0}\Bigl[ 
-\frac{\cos(-\Theta+\delta(2\nu-1))}{2\sin\delta} + \frac{\cos\Theta}{2\delta}\Bigr] = 
-\frac{1}{2}\, (2\nu -1) \, \sin \Theta \ . 
\end{equation}
Noting that 
\begin{equation}
\delta h_{\psi}^{-+}=-\delta y_{\psi}^{-+}=-\lambda \, 
\tilde {\mathcal{E}}_{\phi\psi}^{\psi(-+)}
\end{equation}
we finally get the same result as in (\ref{geometric_dh}).


\section{Third order coefficients in the pure bulk or boundary case}
\setcounter{equation}{0}

In section~2 we explained how to calculate higher order coefficients in
a theory with bulk and boundary perturbations. Actually, the techniques
used there can also be easily generalised to the pure bulk
or pure boundary case; this will be sketched in the following. We 
begin with a discussion  about which coefficients
of the third order terms in the beta functions contain universal 
quantities, paralleling the discussion in section~2.1. In section~4.2 we 
then describe how to obtain
useful formulae for these coefficients.

\subsection{Universal quantities}\label{subsec_4.1}

In this subsection we shall only consider the pure bulk theory; the discussion 
for the pure boundary case is very similar.
Consider a  conformal field theory perturbed by
\begin{equation}\label{deltaS_bulk}
\delta S=\sum_i\lambda^i\,l^{-y_i}\int\!\! d^{2}z\,   \,\phi_i\, (z,\bar{z})\ ,
\end{equation}
where as before $y_l=2-\Delta_l$ are the anomalous dimensions, $l$ is   
a renormalisation length scale, and 
$\lambda_i$ are the dimensionless coupling constants.
 The beta functions have the general form 
\begin{equation}\label{betafunctions_bulk}
\beta^l=y_l\lambda^l+\sum_{ij}\mathcal{C}_{ij}^{l}
\lambda^i\lambda^j+\sum_{ijk}\mathcal{F}_{ijk}^{l}
\lambda^i\lambda^j\lambda^k+\mathcal{O}(\lambda^4) \, .
\end{equation}
We take the constants $\mathcal{C}_{ij}^{l}$ and
$\mathcal{F}_{ijk}^{l}$ to be totally symmetric
in $i,j$ and $i,j,k$, respectively.
Under a general change of scheme the coupling constants are redefined as 
\begin{equation}\label{redef4}
\tilde{\lambda}^l:=\lambda^l+\sum_{ij} c_{ij}^{l}\lambda^i
\lambda^j+\sum_{ijk} f_{ijk}^{l}\lambda^i\lambda^j\lambda^k+
\mathcal{O}(\lambda^4)\ ,
\end{equation}
where the $c_{ij}^{l}$ are, without loss of generality, symmetric in 
$i$ and $j$.  The beta functions in the new scheme are
\begin{equation}\label{change}
\begin{array}{r@{}l@{}l}
\tilde{\beta}^l\,=\,&y_l\tilde{\lambda}^l&+
\displaystyle{\sum_{ij}}\tilde{\lambda}^i\tilde{\lambda}^j
\left(\mathcal{C}_{ij}^{l}+c_{ij}^{l}(y_i+y_j-y_l)\right)\\
&+\displaystyle{\sum_{ijk}}&\tilde{\lambda}^i\tilde{\lambda}^j
\tilde{\lambda}^k\bigg[\mathcal{F}_{ijk}^{l}+
f_{ijk}^{l}(y_i+y_j+y_k-y_l)\\
&&+\frac{1}{3}\sum_m\sum_{\mathrm{perm}(i,j,k)}\left(c_{mi}^{l}\mathcal{C}_{jk}^{m}
-\mathcal{C}_{mi}^{l} c_{jk}^{m}-c_{mi}^{l}
c_{jk}^{m}(y_m+y_i-y_l)\right)\bigg]  +\mathcal{O}(\!\tilde{\lambda}^4)\ .
\end{array}
\end{equation}
We observe that the second order coefficients $\mathcal{C}_{ij}^{l}$
do not change under this transformation if and only if the second order resonance
condition $y_i+y_j=y_l$ is satisfied. 
 As for the coefficients 
$\mathcal{F}_{ijk}^{l}$ at the cubic powers of the 
couplings, it  
 can be seen from (\ref{change}) that the basic requirement for $\mathcal{F}_{ijk}^{l}$  to
be universal is that it satisfies the resonance 
condition $y_i+y_j+y_k=y_l$. However, even if the resonance condition is satisfied, 
the third line in (\ref{change})
shows that the corresponding  coefficient may not be invariant 
under general scheme changes because of the lower order coefficients $\mathcal{C}_{ij}^{l}$. 
The resulting transformations of the resonant coefficients 
are parametrised by the tensors $c_{ij}^{k}$. For an $n$-dimensional coupling 
space the dimension of the space of coefficients $\mathcal{F}_{ijk}^{l}$ is 
$\frac{n^{2}(n+1)(n+2)}{6}$ while that of the coefficients $c_{ij}^{k}$ 
is $\frac{n^{2}(n+1)}{2}$.  Depending on how many coefficients are resonant, 
there may be some functions defined on these resonant coefficients 
which are invariant and thus give universal quantities. For example if all couplings 
are marginal ($y_{i}=0$ for  all $i$) then generically there must be a  subspace 
of scheme independent coefficients of dimension $\frac{n^{2}(n+1)(n-1)}{6}$.   
  
While in general it is hard to write out  explicit expressions for  universal 
quantities in terms of $\mathcal{F}_{ijk}^{l}$ and $\mathcal{C}_{ij}^{l}$ 
we can do so in the absence of second order resonances
because we can then use
a special scheme in which all $\mathcal{C}_{ij}^{l}$ vanish.
Given a cubic resonance $y_i+y_j+y_k=y_l$, the values of the cubic coefficients 
$ \tilde{\mathcal{F}}_{ijk}^{l}$ in that scheme are  universal and  can be 
expressed via the coefficients in an arbitrary scheme as
\begin{equation}\label{change_quadrtozero}
\tilde{\mathcal{F}}_{ijk}^{l} =
\mathcal{F}_{ijk}^{l}+
\frac{1}{3}\sum_{\mathrm{perm}(i,j,k)}\sum_{m}\frac{\mathcal{C}_{ij}^{m}
\mathcal{C}_{mk}^{l}}{y_l-y_k-y_m}\,.
\end{equation}
The scheme independence of  (\ref{change_quadrtozero}) can be checked 
directly using (\ref{redef4}) and (\ref{change}).
\medskip
  
We can also consider  a situation analogous to the one considered in section 2.1 when 
the universal quantity  gives  dimension shifts 
under a truly marginal deformation. Let $\lambda$ be a coupling constant corresponding
to an exactly marginal operator $\phi(z,\bar z)$. 
The beta functions for the other operators $\phi_{l}(z,\bar z)$ have  the form 
\begin{eqnarray}
\beta^l&=& \sum_{k} (y_l\delta_{k}^{l} + \lambda \mathcal{C}_{\phi k}^{l} + \lambda^{2} 
 \mathcal{F}_{\phi\phi k}^{l}  )\lambda^k + 
 \sum_{ij}(\mathcal{C}_{ij}^{l} + \lambda \mathcal{F}_{\phi ij}^{l})
\lambda^i\lambda^j 
 \nonumber \\
&& + \sum_{ijk}\mathcal{F}_{ijk}^{l}\lambda^i\lambda^j\lambda^k +
  \lambda^{2}\mathcal{C}_{\phi\phi}^{l}    +  
\lambda^{3}\mathcal{F}_{\phi\phi\phi}^{l} + \dots \ .
\end{eqnarray}
We can  make the beta functions $\beta^{i}$ homogeneous in $\lambda^{i}$ up
to cubic order by a coupling constant redefinition
\begin{equation}
\tilde \lambda^{i}= \lambda^{i} + \frac{\mathcal{C}_{\phi\phi}^{i}}{y_{i}} \lambda^{2} 
+ \frac{\mathcal{F}_{\phi\phi\phi}^{i}}{y_{i}}\lambda^{3}\ .
\end{equation}
This redefinition is possible because $\lambda$ is truly marginal.
The anomalous dimensions of  the operators $\phi_{i}$ are then 
given by the eigenvalues of the matrix 
\begin{equation} 
D_{i}^{j}(\lambda) \equiv \left( \frac{ \partial \tilde \beta^{j}}{\partial \tilde \lambda^{i}}
\right)_{\tilde \lambda^{k}=0}
\end{equation} 
and have the form
 \begin{equation}
y_{i}[\lambda] = y_{i} + \lambda \delta_{i}^{(1)} + \lambda^{2}\delta_{i}^{(2)} + \dots  \ .
\end{equation} 
A straightforward computation yields 
\begin{equation}
\delta_{i}^{(1)} = \mathcal{C}_{\phi i}^{i} 
\end{equation}
and 
\begin{equation}\label{d2}
\delta_{i}^{(2)}=\mathcal{F}_{\phi\phi i}^{i} - 
2\sum_{k}\frac{\mathcal{C}_{\phi\phi}^{k}\mathcal{C}_{ik}^{i}}{y_{k}} + 
\sum_{k\ne i} \frac{\mathcal{C}_{\phi i}^{k}\mathcal{C}_{\phi k}^{i}}{y_{i}-y_{k}}\ .
\end{equation}
The coefficients $\mathcal{C}_{\phi i}^{i}$ are resonant and thus universal. 
One can also check that (\ref{d2}) is invariant 
under an arbitrary coupling constants redefinition of the form
\begin{equation}
\tilde \lambda^{l}= f^{l}(\lambda) + \sum_{k}f_{k}^{l}(\lambda)\lambda^{k} 
+ \sum_{ik}f_{ik}^{l}(\lambda)\lambda^{i}\lambda^{k} 
+ \sum_{ijk}f_{ijk}^{l}\lambda^{i}\lambda^{j}\lambda^{k}  \ ,
\end{equation}
where $f^{l}$, $f^{l}_{k}$, $f^{l}_{ik}$ are polynomial functions of $\lambda$.

Finally let us mention the well known fact that if there is a single running coupling 
constant whose UV dimension is marginal, then both the quadratic and cubic terms 
in its beta function are universal.

\subsection{Computation of coefficients}

Now that we have understood which coefficients are universal,
we can ask how they can be calculated explicitly. As in the 
bulk-boundary case discussed in section~2, we can either use
a minimal subtraction scheme (see section~2.2) or the
OPE scheme of section~2.3. As before, the minimal
subtraction scheme is conceptually clearer since one does
not need to introduce beta functions for irrelevant fields. 
However, the calculation is somewhat unwieldy in this
scheme, since one has to isolate the divergences in the
UV cut-off $\epsilon$ for finite IR cut-off $L$. 

In the following we shall only consider universal quantities for
which the calculation in either scheme must give the same answer.
Since the OPE scheme is technically simpler, we shall use
it to determine explicit expressions for these coefficients.
We have also checked that our result agrees with what would
have been obtained in the minimal subtraction scheme (as must 
be the case). Moreover for brevity we will focus on the quantity 
(\ref{change_quadrtozero}). It is straightforward to extend our 
results to the dimension shifts (\ref{d2}) and to a cubic term 
in a beta function of a single marginal coupling.

\subsubsection{Resonant bulk coefficients}

In the OPE scheme the RG equations are determined from
the condition that the variation $\epsilon\partial_\epsilon e^{\delta S}$
vanishes in the limit $\epsilon\to 0$. 
As before we regularise the theory by point splitting, {\it i.e.}\ we
introduce a sharp UV cut-off $\epsilon$. In addition we have an
IR cut-off $L$. To cubic order in the couplings we have 
\begin{eqnarray}
e^{\delta S} & =& 1 + \sum_{i}\lambda^{i}\epsilon^{-y_{i}}\int\!\! d^{2} z\, \phi_{i}(z,\bar z) + 
\frac{1}{2!}\sum_{ij}\lambda^{i}\lambda_{j} \epsilon^{-y_{i}-y_{j}}\iint\!\! d^{2}z_{1}d^{2}z_{2}\,
\theta_{12}\phi_{i}(z_{1},\bar z_{1})\phi_{j}( z_2 , \bar z_2 ) \nonumber \\
&& + \frac{1}{3!}\sum_{ijk}\lambda^{i}\lambda^{j}\lambda^{k} \epsilon^{-y_{i}-y_{j}-y_{k}}\!\!
\iiint\!\! d^2 z_1 d^2 z_2 d^2 z_3\, \theta_{12}\theta_{23}\theta_{13} 
\phi_{i}(z_1,\bar z_1)\phi_{j}(z_2, \bar z_2)\phi_{k}(z_3, \bar z_3) + ... \ , \nonumber
\end{eqnarray} 
where $\theta_{ij}=\theta(|z_i-z_j|-\epsilon)\theta(L-|z_i-z_j|)$. The variation 
$\epsilon \partial_{\epsilon}$ 
of this expression can be computed using (\ref{betafunctions_bulk}).  
Setting $\epsilon \partial_{\epsilon}e^{\delta S}\sim 0$ at second order in 
the couplings one obtains the well known expression 
\begin{equation}
\mathcal{C}_{ij}^{m}=\pi\,C_{ij}{}^{m} \ ,
\end{equation}
where $\tri{C}{i}{j}{m}$ are the bulk OPE coefficients 
(\ref{bulkbulkOPE}). 
At the cubic order  we have the equation
\begin{equation}\label{variation_third_order}
\begin{array}{r@{}l}
\displaystyle{0 \esim \lambda^i\lambda^j\lambda^k\!\! \sum_{\mathrm{perm}(i,j,k)}\bigg[}&
\displaystyle{-\frac{1}{2}\epsilon^{-y_i-y_j-y_k+1}
\int d^2z_1d^2z_2d^3z_3\,\delta_{12}^\epsilon\theta_{13}
\theta_{23}\,\phi_i(z_1,\bar{z}_1)\phi_j(z_2,\bar{z}_2)
\phi_k(z_3,\bar{z}_3)}\\[18pt]
&\displaystyle{+\sum_m\pi C_{ij}{}^{m}
\epsilon^{-y_m-y_k}\int d^2z_1d^2z_2\,\theta_{12}\,
\phi_m(z_1,\bar{z}_1)\phi_k(z_2,\bar{z}_2)}\\[18pt]
&\displaystyle{+\sum_l\mathcal{F}_{ijk}^{l}
\epsilon^{-y_l}\int d^2z\,\phi_l(z,\bar{z})\,\bigg]}\ .
\end{array}
\end{equation}
As was discussed in section 2.3 the above equation in general may still have
divergences. However, in the resonant case, {\it i.e.}\ if 
$y_i+y_j+y_k=y_l$ no such complications arise. Then we can write 
\begin{eqnarray}\label{Coefficient_0}
(\mathcal{F}_{ijk}^{l})_{\rm res}&=&\lim_{\epsilon \to 0} \frac{\epsilon}{12}
\int\!\! d^2z_2\,
\theta_{20}\sum_{\mathrm{perm}(ijk)}\,\bigg[\int d^2z_1\,\theta_{10}\, 
\delta_{12}^\epsilon
\cor{\phi_i(z_1,\bar{z}_1)\phi_j(z_2,\bar{z}_2)\phi_k(0)%
\phi_l(\infty)} \nonumber \\[4pt]
&&\qquad-2\pi\sum_m \epsilon^{y_i+y_j-y_m-1}
\tri{C}{i}{j}{m}\cor{\phi_m(z_2,\bar{z}_2)\phi_k(0)\phi_l(\infty)}
\bigg]\ , \label{q1}
\end{eqnarray}
where $\theta_{10}=\theta(|z_1|-\epsilon)\theta(L-|z_1|)$, and similarly for $\delta_{10}$.
As before, we consider spinless fields, for which we can 
express the four-point correlator in terms of conformal blocks,
\begin{equation}
\cor{\phi_1(z_1,\bar{z}_1)\phi_2(z_2,\bar{z}_2)\phi_3(z_3,\bar{z}_3)
\phi_4(z_4,\bar{z}_4)} = \prod_{i<j} |z_{ij}|^{2 (\delta-h_i-h_j)}
Y_{1234}(\eta,\bar{\eta})
\end{equation}
with
\begin{equation}
z_{ij}=z_i-z_j\ ,\qquad 
\delta=\frac{1}{3}\sum_{i=1}^{4} h_i\ , \qquad
\eta=\frac{z_{12}z_{34}}{z_{13}z_{24}}\ ,
\end{equation}
and
\begin{equation}\label{bulk_conformal_block}
Y_{1234}(\eta,\bar{\eta})=\left\{
\begin{array}{l}
\sum_mC_{12}{}^{m}C_{m3}{}^{4}F^m_{12,34}(\eta)\tilde{F}^m_{12,34}(\bar{\eta})\\[4pt]
\sum_mC_{32}{}^{m}C_{m1}{}^{4}F^m_{32,14}(1-\eta)\tilde{F}^m_{32,14}(1-\bar{\eta})\\[4pt]
\sum_mC_{13}{}^{m}C_{m2}{}^{4}F^m_{13,24}(1/\eta)\tilde{F}^m_{13,24}(1/\bar{\eta})\ .
\end{array}
\right.
\end{equation}
The conformal blocks are normalised such that
\begin{equation}\label{conf_block_asymptotic}
F^m_{12,34}(\eta)\sim\eta^{h_m-\delta} \qquad 
\tilde{F}^m_{12,34}(\bar{\eta}) \sim \bar{\eta}^{{h}_m-\delta} \qquad 
\hbox{for $\eta,\bar{\eta}\rightarrow 0$.}
\end{equation}
Replacing the variables in the first part of the square 
brackets in (\ref{Coefficient_0}) by an angular variable $\varphi$ 
and the cross ratio $\eta$,
\[
z_1=z_2+\epsilon e^{i\varphi} \ , \qquad
z_2=\epsilon \, e^{i\varphi}\, \frac{1-\eta}{\eta}\ ,
\]
the angular variable can be integrated out by the integral over $z_1$,
and one finds
\begin{eqnarray}\label{third_order_1}
\displaystyle{(\mathcal{F}_{ijk}^{l})_{\rm res}}&= & 
\displaystyle{\lim_{\epsilon\rightarrow 0}\frac{\pi}{6}\int d^2\eta
\sum_{\mathrm{perm}(i,j,k)}\theta(1-|\eta|)
\theta(|\eta|-\tfrac{\epsilon}{L}) \bigg\{\theta(\tfrac{1}{2}-\re\eta)
\theta(|\eta|^2(\tfrac{L^2}{\epsilon^2}-1)+2\re\eta-1)
} \nonumber  \\
& &\qquad \qquad \qquad\qquad \displaystyle{\times
|\eta|^{-y_i-y_j-2y_k-4\delta+4}|1-\eta|^{2\delta+y_j+y_k-4}
Y_{ij,kl}(\eta,\bar{\eta})} \\
&& \displaystyle{\qquad\qquad \qquad \qquad \qquad -\sum_m C_{ij}{}^{m}
C_{mk}{}^{l}|\eta|^{y_l-y_m-y_k-2}\bigg\}\ .} \nonumber
\end{eqnarray}
In the second term in the bracket of (\ref{q1}), we changed variables to
$z_2=\epsilon/\eta$. The function in the last line in (\ref{third_order_1}) can be 
integrated explicitly, and we obtain 
\begin{eqnarray}\label{third_order_2}
\displaystyle{(\mathcal{F}_{ijk}^{l})_{\rm res}} \hspace*{-0.1cm}
& = & \hspace*{-0.1cm}
\displaystyle{\lim_{\epsilon\rightarrow 0}\sum_{\mathrm{perm}(i,j,k)}\bigg[
\frac{\pi}{6}\int d^2\eta\,\theta(1-|\eta|)
\theta(|\eta|-\tfrac{\epsilon}{L})\theta(\tfrac{1}{2}-\re\eta)} \nonumber \\
& &  \hspace*{-0.1cm} \displaystyle{
|\eta|^{-y_i-y_j-2y_k-4\Delta+4}|1-\eta|^{2\Delta+y_j+y_k-4}
Y_{ij,kl}(\eta,\bar{\eta})-\frac{\pi^2}{3}\sum_m\frac{C_{ij}{}^{m}
C_{mk}{}^{l}}{y_l-y_k-y_m}\left(\frac{\epsilon}{L}
\right)^{y_l-y_k-y_m}\bigg]} \nonumber \\[5pt]
& &\displaystyle{\qquad+\frac{\pi^2}{3}\sum_{\mathrm{perm}(i,j,k)}
\sum_m\frac{C_{ij}{}^{m}
C_{mk}{}^{l}}{y_l-y_k-y_m}\ .}
\end{eqnarray}
The  universal quantity 
$\tilde{\mathcal{F}}_{ijk}^{l}$  defined in (\ref{change_quadrtozero}) 
is then simply given by the first two lines of (\ref{third_order_2}).

It can be checked using the asymptotics (\ref{conf_block_asymptotic}) 
and the properties of conformal blocks (\ref{bulk_conformal_block}) that the integral 
(\ref{third_order_1}) converges in the regions $\eta \sim \epsilon/L \to 0$, 
$|1-\eta| \sim  \epsilon/L \to 0$ and $|\eta|\sim L/\epsilon \to \infty$. One can thus safely set 
$\epsilon=0$ in (\ref{third_order_1}) to obtain an integral expression 
\begin{eqnarray}\label{third_order_3}
\displaystyle{(\mathcal{F}_{ijk}^{l})_{\rm res}} & =& 
\displaystyle{
\frac{\pi}{6}\int d^2\eta\sum_{\mathrm{perm}(i,j,k)}
\theta(1-|\eta|)\theta(\tfrac{1}{2}-\re\eta)\bigg\{}\displaystyle{
\,|\eta|^{2r+y_i+y_j-4}|1-\eta|^{2r+y_j+y_k-4}
Y_{ij,kl}(\eta,\bar{\eta})} \nonumber \\[5pt]
& &\qquad \qquad \qquad \qquad \qquad  \qquad \qquad \qquad \qquad
\displaystyle{-\sum_m C_{ij}{}^{m}
C_{mk}{}^{l}|\eta|^{y_i+y_j-y_m-2}\bigg\}}\ .
\end{eqnarray}
 Note also that expression (\ref{third_order_3}) is $L$ independent. In particular this means  that 
 the infrared divergences that were present in individual summands in 
(\ref{third_order_2}) mutually cancel each other.  This agrees with the general results of 
\cite{GM1,GM2}.\footnote{The perturbation expansion for Wilson coefficients 
proposed in \cite{GM1,GM2} was shown to be IR finite to all orders 
under certain assumptions on the 
UV renormalisation scheme. 
As we are interested in scheme independent quantities their result applies.}

By suitable changes of the integration variable $\eta$ in the
terms with permuted indices $i,j,k$ it is possible to write
$\mathcal{F}_{ijk}^{l}$ by means of integrals over
three disjoint subsets  tiling 
 the whole $\eta$-plane.  
Consider  the
transformation $\eta\mapsto 1-\eta$, for which the cut-off
functions in the integral (\ref{third_order_3}) become 
$\theta(1-|\eta-1|)\theta(\re\eta-\tfrac{1}{2})$.
The asymptotics (\ref{bulk_conformal_block}) for
$Y_{ij,kl}$ are such that the divergence of the transformed integrand
that arise from the limit $\eta\mapsto 1$ is again canceled, once
we take the transformed subtractions, {\it i.e.}\ the second line in
(\ref{third_order_3}), into account.
The other transformation is $\eta\mapsto1/\eta$. In this case the cut-off 
functions read $\theta(|\eta|-1)\theta(|\eta-1|-1)$ after the 
transformation, and the divergence
of the corresponding integrand for $\eta\rightarrow \infty$ is 
canceled as well. Together, the regions carved out by the 
cut-off functions for the three coordinate choices tile the 
whole $\eta$ plane. Using this we 
 can recast (\ref{third_order_3}) as 
\begin{equation}
(\mathcal{F}_{ijk}^{l})_{\rm res}\,=\,
\frac{\pi}{3}\int\!\! d^2\eta\,   \Bigl[
\,|\eta|^{2r+y_i+y_j-4}|1-\eta|^{2r+y_j+y_k-4}
Y_{ij,kl}(\eta,\bar{\eta}) - S_{ijkl}(\eta) \Bigr] \ ,
\end{equation}
where 
\begin{eqnarray}
S_{ijkl}(\eta) & =&  \sum_m C_{ij}{}^{m} C_{mk}{}^{l}
\, |\eta|^{y_i+y_j-y_m-2}\theta(1-|\eta|)\theta(\tfrac{1}{2}-\re\eta) 
\nonumber \\
&&+ 
\sum_m C_{kj}{}^{m} C_{mi}{}^{l}
\, |1-\eta|^{y_k+y_j-y_m-2}\theta(1-|\eta-1|)\theta(\re\eta-\tfrac{1}{2}) \nonumber 
\\ && +
 \sum_m C_{ik}{}^{m} C_{mj}{}^{l}
 \, |\eta|^{-y_i-y_k+y_m-2}\theta(|\eta|-1)\theta(|\eta-1|-1)\ .
\end{eqnarray}
In this form the integration runs over the whole $\eta$-plane. Although the subtraction function $S_{ijkl}(\eta)$ 
still has a piecewise form it is expressed quite explicitly.

\subsubsection{Resonant boundary coefficients}

On the boundary the computation can be done in a similar way as in the bulk.
We consider a boundary perturbation of the form
\begin{equation}
\delta S=\sum_s\mu^s\,\epsilon^{-y_s}\int dx\,\psi_s(x)\ ,
\end{equation}
where now $y_s=1-h_s$. Up to the third order in the couplings
the RG equations take the form
\begin{equation}
\dot{\mu}^s=y_s\mu^s+\sum_{p,q}\mathcal{D}_{pq}^{s}
\mu^p\mu^q+\sum_{p,q,r}\mathcal{G}_{pqr}^{s}\mu^p
\mu^q\mu^r+\ldots\ .
\end{equation}
As before, we only introduce counterterms at the quadratic order for 
marginal or relevant fields, and the corresponding coefficients are 
\begin{equation}
\mathcal{D}_{pq}^{s}=D_{pq}{}^{s}\ ,
\end{equation}
where $D_{pq}{}^{s}$ is the OPE coefficient of two boundary fields
(\ref{boundaryboundaryOPE}).
In the resonant case where we have $y_s=y_p+y_q+y_r$, the 
coefficient $(\mathcal{G}_{pqr}^{s})_{\rm res}$
can be written as
\begin{equation}\label{third_order_coeff_boundary_0}
\begin{array}{r@{}l}
(\mathcal{G}_{pqr}^{s})_{\rm res} \,=\,
\displaystyle{\frac{1}{6}\lim_{\epsilon\to 0}\sum_{\mathrm{perm}(p,q,r)}
\bigg\{ }&\displaystyle{\epsilon \int_{2\epsilon}^{L}\cor{
\psi_p(0)\psi_q(\epsilon)\psi_r(x)\psi_s(\infty)} }\\
&\displaystyle{+\epsilon \int_{-L}^{-\epsilon}\cor{
\psi_p(x)\psi_q(0)\psi_r(\epsilon)\psi_s(\infty)} }\\
&\displaystyle{-\sum_t\epsilon^{-y_t-y_r+y_s}D_{pq}{}^{t}\int_{\epsilon}^{L}\cor{
\psi_t(0)\psi_r(x)\psi_s(\infty)} }\\
&\displaystyle{-\sum_t\epsilon^{-y_t-y_p+y_s}D_{qr}{}^{t}\int_{-L}^{-\epsilon}\cor{
\psi_p(x)\psi_t(0)\psi_s(\infty)} \bigg\}}\ .
\end{array}
\end{equation}
By similar arguments as in the previous subsection we can find
\begin{equation}\label{third_order_coeff_boundary}
\begin{array}{r@{}l}
(\mathcal{G}_{pqr}^{s})_{\rm res}\,=\,
\displaystyle{\frac{1}{6}\int_0^1 d\eta\sum_{\mathrm{perm}(p,q,r)}
}&\displaystyle{\bigg\{ \eta^{r+y_p+y_q-2}(1-\eta)^{r+y_q+y_r-2}\,Y_{pq,rs}(\eta)}\\
&\displaystyle{-\sum_t D_{pq}{}^{t}D_{tr}{}^{s}\eta^{y_p+y_q-y_t-1}
-\sum_t D_{qr}{}^{t}D_{pt}{}^{s}(1-\eta)^{y_q+y_r-y_t-1}\bigg\} }\ ,
\end{array}
\end{equation}
where
\begin{equation}
Y_{pq,rs}(\eta)=\sum_{t}D_{pq}{}^{t}D_{tr}{}^{s}F_{pq,rs}^t(\eta)=
\sum_tD_{qr}{}^{t}D_{pt}{}^{s}F_{sp,qr}^t(1-\eta)\ .
\end{equation}
Here the conformal blocks $F_{pq,rs}^t(\eta)$ have cuts running from
$-\infty$ to zero, and from $1$ to $+\infty$. In addition, their 
asymptotic behaviour is 
\begin{equation}
F_{pq,rs}^t(\eta)\sim\eta^{h_t-\delta}\qquad(\eta\rightarrow 0)\ ,
\end{equation}
where $\delta$ is defined as before, {\it i.e.}\ 
$\delta=\tfrac{1}{3}(h_p+h_q+h_r+h_s)$. 
Finally, the scheme-independent quantity is given by
\begin{equation}
\tilde{\mathcal{G}}_{pqr}^{s}=
(\mathcal{G}_{pqr}^{s})_{\rm res}
+\frac{1}{6}\sum_{\mathrm{perm}(p,q,r)}
\sum_t\frac{D_{pq}{}^{t}(D_{tr}{}^{s}+D_{rt}{}^{s})}{y_s-y_r-y_t} \ .
\end{equation}

In the case where several irreducible boundary conditions are involved, one has
to keep track of their labels, and bear in mind the superselection rules, in particular
the order of operators. This leads to additional splittings and recombinations of the 
integrals over four-point functions and subtractions. Apart from this 
technicality, it is however straightforward to include the boundary labels. 
We have refrained from writing them explicitly to keep the formulae simpler.

\section{Conclusions}

In this paper we have studied conformal perturbation theory beyond
the leading order. We have shown that, at least up to quadratic order,
the combined bulk boundary perturbation problem is renormalisable,
using the minimal subtraction scheme. We also discussed the more
commonly used `Wilsonian' OPE scheme, and found it to have some
shortcomings at higher order in perturbation theory. We identified
systematically the universal (scheme-independent) quantities, and 
gave explicit formulae for them at third order in terms of integrals of 
conformal 4-point functions. Finally, we explained how essentially
the same analysis works for the pure bulk and pure boundary case.
It seems plausible that similar techniques should allow one to
prove renormalisability at arbitrary order in perturbation theory, but 
we have not attempted to do so. 

Our work was originally motivated by the question of how the 
dependence of the conformal dimension of a boundary changing field 
upon a bulk modulus can be understood from the world-sheet 
perspective. Our considerations demonstrate that this effect is 
captured
by a certain universal quadratic RG coefficient, for which we gave
an explicit formula. This result should also have interesting 
applications in other contexts; in particular, it provides a 
world-sheet method to study the stability of brane setups under 
arbitrary bulk deformations.

\section*{Acknowledgements} 

The research of MRG and CSC has been partially supported by 
the Swiss National Science Foundation and the Marie Curie network
`Constituents, Fundamental Forces and Symmetries of the Universe'
(MRTN-CT-2004-005104). CSC thanks the University of Heriot-Watt,
Edinburgh, for hospitality, where part of this work was carried out. 
We thank Stefan Hohenegger, Christoph Keller, and Andreas Ludwig 
for useful conversations.

\end{document}